\documentclass[floats,floatfix,showpacs,amssymb,prd,twocolumn,superscriptaddress,nofootinbib,nolongbibliography,reprint]{revtex4-1}

\usepackage{amssymb,amsmath,verbatim,mathtools,needspace,enumitem,etoolbox,graphicx,physics,microtype,afterpage,xspace,tabularx,lmodern,multirow}
\usepackage{gensymb}
\usepackage[dvipsnames, usenames]{xcolor}
\definecolor{linkcolor}{rgb}{0.0,0.3,0.5}
\usepackage[unicode, colorlinks=true, linkcolor=linkcolor, citecolor=linkcolor, filecolor=linkcolor, urlcolor=linkcolor, linktocpage, breaklinks]{hyperref}
\usepackage[all]{hypcap}
\usepackage[T1]{fontenc}
\usepackage[utf8]{inputenc}
\usepackage[usenames,dvipsnames]{xcolor}
\hypersetup{colorlinks=true,citecolor=romared,linkcolor=romared,urlcolor=romared}

\definecolor{romared}{RGB}{142,0,28}

\newcommand{\be}{\begin{equation}}
\newcommand{\ee}{\end{equation}}

\def\be{\begin{equation}}
\def\ee{\end{equation}}
\newcommand{\beq}{\begin{eqnarray}}
\newcommand{\eeq}{\end{eqnarray}}

\usepackage{makecell}
\usepackage{soul}

\begin{document}

\title{Dynamical friction from scalar dark matter in the relativistic regime}
\author{Dina Traykova}
\email{dina.traykova@physics.ox.ac.uk}
\affiliation{Astrophysics, University of Oxford, DWB, Keble Road, Oxford OX1 3RH, UK}
\author{Katy Clough}
\email{katy.clough@physics.ox.ac.uk}
\affiliation{Astrophysics, University of Oxford, DWB, Keble Road, Oxford OX1 3RH, UK}
\author{Thomas Helfer}
\email{thelfer1@jhu.edu}
\affiliation{Department of Physics and Astronomy, Johns Hopkins University,
3400 N. Charles Street, Baltimore, Maryland, 21218, USA}
\author{Emanuele Berti}
\email{berti@jhu.edu}
\affiliation{Department of Physics and Astronomy, Johns Hopkins University,
3400 N. Charles Street, Baltimore, Maryland, 21218, USA}
\author{Pedro G. Ferreira}
\email{pedro.ferreira@physics.ox.ac.uk}
\affiliation{Astrophysics, University of Oxford, DWB, Keble Road, Oxford OX1 3RH, UK}
\author{Lam Hui}
\email{lh399@columbia.edu}
\affiliation{Center for Theoretical Physics, Department of Physics, Columbia University, New York, NY 10027, USA}

\date{Received \today; published -- 00, 0000}

\begin{abstract}
Light bosonic scalars (e.g. axions) may form clouds around black holes via superradiant instabilities, or via accretion if they form some component of the dark matter. It has been suggested that their presence may lead to a distinctive dephasing of the gravitational wave signal when a small compact object spirals into a larger black hole. Motivated by this, we study numerically the dynamical friction force on a black hole moving at relativistic velocities in a background scalar field with an asymptotically homogeneous energy density. We show that the relativistic scaling is analogous to that found for supersonic collisional fluids, assuming an approximate expression for the pressure correction which depends on the velocity and scalar mass. 
While we focus on a complex scalar field, our results confirm the expectation that real scalars would exert a force which oscillates between positive and negative values in time with a frequency set by the scalar mass. The complex field describes the time averaged value of this force, but in a real scalar the rapid force oscillations could in principle leave an imprint on the trajectory.
The approximation we obtain can be used to inform estimates of dephasing in the final stages of an extreme mass ratio inspiral.

\end{abstract}
\keywords{Black holes, Perturbations, Gravitational Waves, Horndeski, Scalar Tensor, dark matter}

\maketitle


\section{Introduction}

Black holes (BHs) provide us with a unique laboratory to study matter in strong gravity conditions. The interplay of new forms of matter with BHs can help us shed light on the matter composition and properties. Such investigations are particularly timely given the recent imaging of a BH at the centre of the galaxy M87 \cite{Akiyama:2019cqa}, ongoing detections of gravitational wave (GW) signals from the collisions of BH binary systems by the Advanced LIGO/Advanced Virgo/KAGRA network \cite{TheLIGOScientific:2016pea, TheVirgo:2014hva, Somiya:2011np, Aasi:2013wya}, and the promise of future detections by third-generation ground-based detectors \cite{Punturo:2010zz, Reitze:2019iox, Saleem:2021iwi} and space based-detectors like LISA, Tianqin and Taiji \cite{Audley:2017drz, Hu:2017mde, Luo:2015ght, Barausse:2020rsu}.

Light bosonic scalar particles, such as axions, provide a well-motivated extension of Standard Model physics \cite{Peccei:1977hh, Arvanitaki:2009fg} (see \cite{Peccei:2006as, Marsh:2015xka} for reviews). 
They are a potential candidate for dark matter (DM) \cite{Preskill:1982cy,Abbott:1982af,Dine:1982ah,Hu:2000ke, Hui:2016ltb} (see \cite{Ferreira:2020fam, Niemeyer:2019aqm, Hui:2021tkt} for reviews), and in such scenarios their accretion onto BHs may result in the formation of distinctive structures \cite{Hui:2019aqm, Clough:2019jpm, Bamber:2020bpu}. 
Furthermore, superradiant instabilities can give rise to gravitationally bound clouds of axions around spinning BHs by amplifying small fluctuations in the field \cite{Detweiler:1980uk, Cardoso:2005vk,  Dolan:2007mj, Herdeiro:2014goa, Arvanitaki:2014wva, Arvanitaki:2010sy},
in which case they need not be a major component of the DM (see \cite{Brito:2015oca} for a review).

In the case of superradiant clouds, it has been proposed that the GW signal from a smaller compact object orbiting the BH, passing through the bosonic environment, may carry distinctive signatures of its presence, detectable via LISA observations of EMRI signals \cite{Macedo:2013qea, Ferreira:2017pth, Hannuksela:2018izj, Zhang:2019eid}. 
In Ref.~\cite{Hannuksela:2018izj}, such effects were contrasted with the impact of dark matter spikes that may cause a similar dephasing \cite{Eda:2013gg, Eda:2014kra, Kavanagh:2020cfn,  Yue:2019ozq, Bertone:2019irm, Hannuksela:2019vip, Edwards:2019tzf, Yue:2017iwc}, and found to be distinguishable. 
Whilst Ref.~\cite{Hannuksela:2018izj} considered dephasing of the orbit due to the gravitational field of the cloud, dynamical friction should also play an important role, and potentially enhance the effect of the cloud on the evolution of the binary. 
In this article we seek to expand existing results for dynamical friction in light scalar DM to the relativistic regime, which would be relevant for the final stages of inspiral. 

Dynamical friction, first studied by Chandrasekhar \cite{Chandrasekhar:1943ys}, is a drag force experienced by an astrophysical object moving in a fluid or bath of heavy particles. 
The gravitational attraction of the larger object creates an overdensity behind it (a ``gravitational wake''), resulting in a drag force with a form that depends on the nature of the fluid and the perturber. 
In the nonrelativistic limit the dynamical friction force scales with velocity as
\begin{equation}
    F_{\rm d, nonrel} \sim \frac{\log(v)}{v^2}\,.
    \label{eqn:Fd_prop}
\end{equation}
This scaling does not hold in the relativistic regime where $v \sim 1$. (This scaling also breaks down as $v\to0$, where as expected $F_{\rm d, nonrel}(v)\to 0$).
The case of a single point object immersed in scalar matter was studied in the nonrelativistic limit in Ref.~\cite{Hui:2016ltb}, where it was shown that the dynamical friction force is suppressed for light scalar matter. This happens when their wave-like behaviour becomes relevant on the scale of the accretion radius $\sim GM/v^2$ where $M$ is the mass of the object. 

\begin{figure*}[t]
	\centering
	\includegraphics[width=0.8\textwidth]{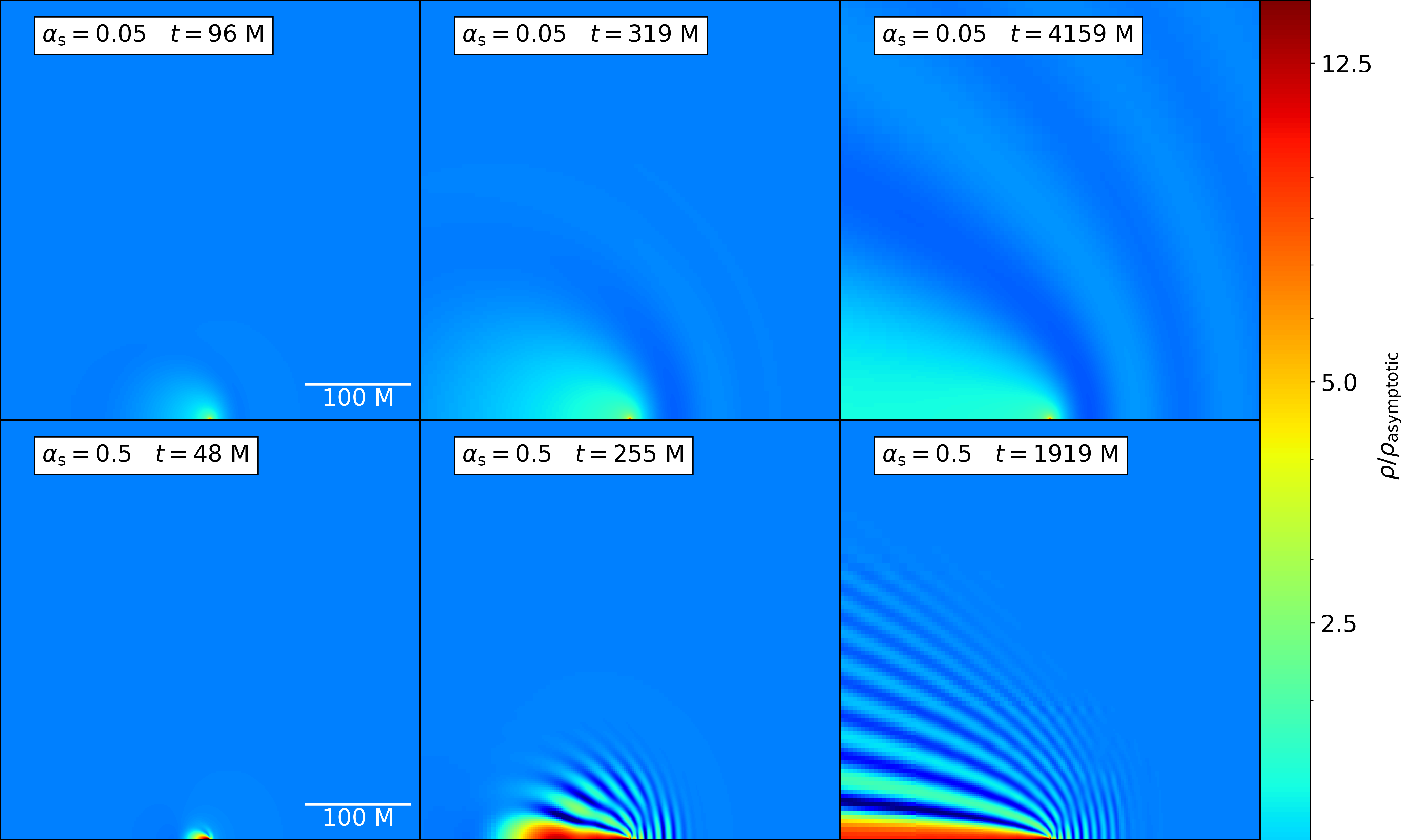}
	\caption{Image illustrating our numerical setup: the density distribution, $\rho$, of the scalar cloud (normalised to the asymptotic value of the density, $\rho_{\rm asymptotic}$) around a BH that is boosted relative to the matter rest frame.
	The two rows show different scalar field masses, $\alpha_{\rm s} = 0.05$ and $0.5$, from top to bottom, increasing in time from left to right (the full time evolution for these simulations can be found at \url{https://youtu.be/5ZDfSwW9NHA} and \url{https://youtu.be/XUACRBfZX-w} for $\alpha_{\rm s} = 0.05$ and $0.5$ respectively). This illustrates the known result that for smaller scalar field masses the tail is less dense and dynamical friction is suppressed.}
	\label{fig:density_mass}
\end{figure*}

\begin{figure*}[t]
	\centering
	\includegraphics[width=0.8\textwidth]{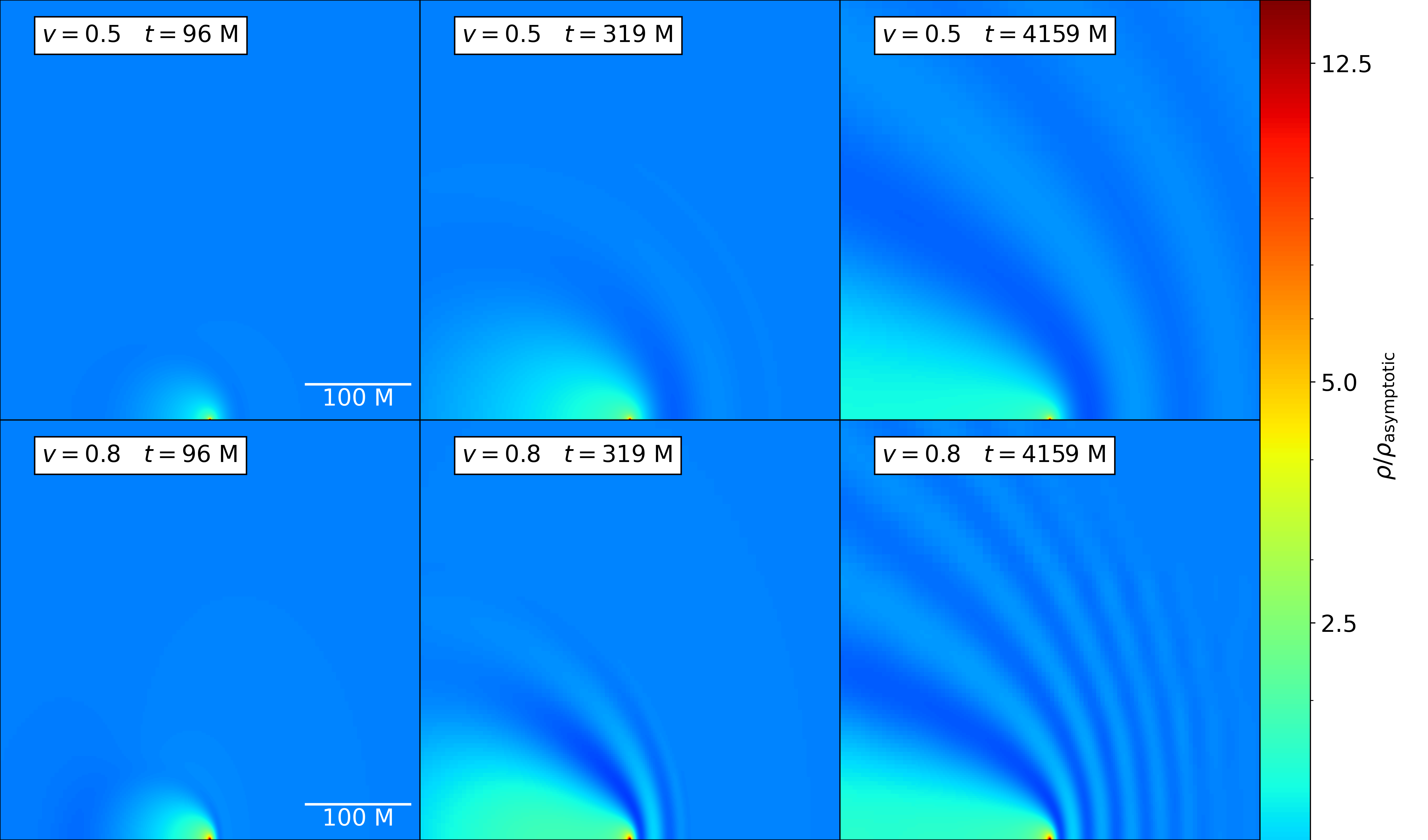}
	\caption{The time evolution (from left to right) of the density profiles of the cloud (normalised to the asymptotic value of the density, $\rho_{\rm asymptotic}$) for two different velocities, $v=0.5$ and $0.8$, and the same scalar mass, $\alpha_{\rm s} = 0.05$.
	This illustrates the change in the profile from the non relativistic to the relativistic regime. We find that, contrary to the nonrelativistic scaling, the dynamical friction force
	does not decrease with velocity, with the density of matter in the cloud slightly higher at $v=0.8$ than $v=0.5$.
	}
	\label{fig:density_v}
\end{figure*}

In this work we investigate the corrections to the scaling of the dynamical friction force with velocity for scalar matter in the relativistic regime, following similar methods to those used by Petrich et al.~\cite{1989ApJ...336..313P} for collisional fluids, where the fluid is numerically evolved to a stationary state from which the force can be measured.

Our setup is illustrated in Fig.~\ref{fig:density_mass},
with examples of the time evolution (increasing from left to right) of the density patterns around the BH for different scalar masses.
We present two of the cases studied in this work: $\alpha_{\rm s}=0.05$ (on the top), which covers the wave-like small mass regime, and $\alpha_{\rm s}=0.5$ (on the bottom), where the wave-like effects begin to be suppressed.
As in Ref.~\cite{Hui:2016ltb}, we can see that larger scalar masses result in higher densities in the stationary cloud that forms behind the BH, leading to a larger dynamical friction force.
Unlike Ref.~\cite{1989ApJ...336..313P}, we carry out our simulations in the rest frame of the scalar matter, such that it has asymptotically zero momentum and a constant energy density far from the BH. 
The BH is boosted relative to the scalar fluid with relativistic velocity $v \sim 0.3-0.8$, and we measure the force experienced by the BH as 
$F_d = dP/dt$ in that frame, which can be shown to be equal to the proper force $dP/d\tau$ measured in the rest frame of the BH  (see Sec. \ref{sec:bgmetric}). This facilitates comparison to known results which tend to be quoted in the rest frame of the perturber.
Fig.~\ref{fig:density_v} illustrates the time evolution (left to right) of the density profiles that form behind the BH for two different velocities: $v=0.5$ on the top and $v=0.8$ on the bottom.
One can see that the overall amplitude of the density and the spread of the tail is dependent on the velocity, and that the force does not decrease significantly as one might expect from the nonrelativistic scaling in Eq.~\eqref{eqn:Fd_prop}. It is in fact larger for the higher velocity case, where the density has a slightly larger amplitude in the cloud behind the BH.

For collisional fluids, Refs.~\cite{1989ApJ...336..313P} and \cite{Barausse:2007ph} showed that there is an overall relativistic correction factor related to the increased effective mass of the perturber at relativistic velocities, plus an adjustment to account for the pressure of the fluid, and we find a similar trend.
However, since inhomogeneous scalar dark matter does not have a well-defined equation of state, we approximate the relation between density and pressure by assuming a plane-wave solution, which makes the pressure contribution velocity-dependent, unlike in normal fluids. The plane wave description holds far from the BH, but as shown in Fig.~\ref{fig:density_v}, the solution close to the BH will be more complex. Nevertheless, using this approximation, we find good agreement with an expression of the form
\begin{equation}
    F_{\rm d} = F_{\rm d, nonrel} \times \gamma^2 (1+v^2)^2 \times \left(1 + \kappa \frac{v^2}{1+v^2}\right)\,,
    \label{eq-dfrel}
\end{equation}
where $\gamma = 1/\sqrt{1-v^2}$ and $\kappa = \kappa(\alpha_{\rm s})\in(0,1)$ characterises the dependence of the pressure correction on the scalar mass, with an $O(1)$ value in the limit of small masses. For larger masses where the wave-like behaviour is no longer significant on the scale of the BH we expect the pressure correction to reduce to zero, ie, $\kappa \sim 0$. The $\kappa$ values obtained in the intermediate mass range studied here can be fit with an approximate linear dependence, as shown in Fig. \ref{fig:kappa_vs_alphas}.

We structure the paper as follows. 
In Sec.~\ref{sec:framework} we describe in detail the physical set up, our numerical implementation, and how the relativistic force is measured. 
In Sec.~\ref{sec:df_bg} we summarize previous relevant studies of dynamical friction and use them to justify our approximation.
In Sec.~\ref{sec:results} we present our simulations, confirming the scaling of the dynamical friction force for a range of scalar masses. 
In Sec.~\ref{Discussion} we discuss our results and propose applications and areas for further work.

For the remainder of the paper we use geometrical units in which $G=c=1$, with the scalar field mass parametrized by the inverse length scale $\mu$. 
One can relate this to the scalar mass $m$ via $\mu = m c / \hbar$. 
We vary the dimensionless quantity $\alpha_{\rm s} = M\mu$, which sets the mass of the BH if the mass of the particle is specified, and vice versa. 
As an example, for the case $\alpha_{\rm s}=1$, this could correspond to a solar mass BH and a scalar with mass $m \sim 10^{-10}$~eV, or a supermassive BH of $M= 10^{10}M_\odot$ with a scalar mass $m \sim 10^{-20}$~eV.
In all of our simulations we set $M=1$.

\section{Framework}
\label{sec:framework}

We work in the decoupling limit in which the backreaction of the scalar field onto the metric is neglected, such that the dynamical friction force is calculated to first order in the density of the field as the rate of change of momentum of the BH spacetime $F_i = \partial_t P_i$. 

Therefore our numerical system is composed of a fixed boosted BH metric, and a complex scalar field that evolves dynamically on this background according to the relativistic Klein-Gordon equation. We describe these components, the force calculation and our numerical implementation in more detail in this section.

\subsection{Background metric}
\label{sec:bgmetric}

We choose to work in the rest frame of the scalar fluid, which allows us to implement spatially homogeneous boundary conditions at large distance from the BH. In this frame the BH is moving with velocity $v$, and has ADM momentum in the $x$ direction $P_x^{\rm ADM} = \gamma M v$, where $\gamma = 1/\sqrt{1-v^2}$ is the usual boost factor.

To obtain the appropriate coordinates we boost the Schwarzschild metric in isotropic coordinates $(\bar t, \bar x, \bar y, \bar z)$ by a factor of $\gamma$ in the $x$ direction. To make the metric time-independent (effectively to prevent the BH from moving across the grid), we add a spatially constant shift to fix the $x$ coordinate to the centre of the BH. We thus obtain our simulation coordinates $(t, x, y, z)$ as
\begin{equation}
    \bar t = (t / \gamma - \gamma v x) \quad 
    \bar x = \gamma x \quad
    \bar y = y \quad \bar z = z ~.
    \label{eqn:coords}
\end{equation}
The $3+1$ ADM decomposition of the Schwarzschild metric in these coordinates is
\begin{equation}
ds^2=-\alpha^2\,dt^2+\gamma_{ij}(dx^i + \beta^i\,dt)(dx^j + \beta^j\,dt)\,,
\end{equation}
where $x^i=x_i$ are the Cartesian coordinates on the grid, and the lapse, shift and nonzero components of the spatial metric are, respectively, 
\begin{equation}
\begin{aligned}
\alpha^2 &= \frac{AB}{\gamma^2(B - A v^2)}\,,\qquad \beta_i = \delta_{ix} A v\,, \\ 
\gamma_{xx} &=  \gamma^2(B - Av^2)\,,
\qquad \gamma_{yy} = \gamma_{zz} =  B \,,
\end{aligned}
\end{equation}
where $$A = \left(\frac{1-M/2\bar r}{1+M/2\bar r}\right)^2\qquad \text{and}\qquad B = \left(1+ \frac{M}{2 \bar r}\right)^4\,,$$ with $\bar r^2 = \gamma^2 x^2 + y^2 + z^2$.
Noting that in these coordinates $\partial_t \gamma_{ij}=0$, the extrinsic curvature $K_{ij}$ can be obtained from its definition, as
\begin{equation}
\begin{aligned}
2 \alpha K_{ij} =& D_i \beta_j + D_j \beta_i\,.
\end{aligned}
\end{equation}
Since the BH is not moving in our chosen frame, we find that it has momentum $P^x=0$, but $P_x = \gamma M v$, the latter quantity reflecting the fact that the normal observers are in the rest frame of the fluid.

Most of the literature to date calculates the dynamical friction force in the rest frame of the BH, since this is analytically more convenient. We therefore need to take care of the frame dependence of the measured force.

Consider the coordinate transformations in Eq.~\eqref{eqn:coords} and their inverse:
\begin{equation}
    t = \gamma (\bar t + v \bar x) \quad 
    x =  \bar x / \gamma \quad
    y = \bar y \quad z = \bar z \,.
\end{equation}

If an object experiences a change in four-momentum $dP_{\bar \mu}$ in the unboosted frame, the transformation to our frame is:
\begin{equation}
    dP_\mu = \frac{\partial x^{\bar\mu}}{\partial x ^\mu} dP_{\bar \mu}\,.
\end{equation}
Thus
\begin{equation}
    dP_x = \gamma dP_{\bar x}
\end{equation}
(for small $dP_{\bar x}$, we have $dP_{\bar t} \sim 0$). The proper time elapsed during this change is $d\tau = d \bar t = dt / \gamma$.
Thus
\begin{equation}
    \frac{dP_x}{dt} = \frac{dP_{\bar x}}{d\bar t}\,.
\end{equation}
This implies that the result for the (coordinate) time derivative of the momentum is the same in both frames.

In terms of the proper time derivative (i.e., the acceleration four-vector) we have
\begin{equation}
    \frac{dP_x}{d\tau} = \gamma \frac{dP_{\bar x}}{d\tau}\,,
\end{equation}
as expected (but note that ours is not a simple Lorentz transformation, differing also by a 
coordinate shift).

\subsection{Scalar field evolution}

We simulate a complex scalar $\varphi$ minimally coupled to gravity, with action
\begin{equation}
	S = \int d^4 x\sqrt{-g} \left[\frac{1}{2}g^{\mu\nu}\nabla_\mu\varphi^*\nabla_\nu\varphi - V(|\varphi|)\right]\,,
\end{equation}
where $V(|\varphi|) = \frac{1}{2}\mu^2|\varphi|^2$.
We solve the second-order Klein-Gordon equation for the real and imaginary parts of the complex scalar field by decomposing them into two coupled first-order equations. The real part and the imaginary part both then obey
\begin{align}
\partial_t \varphi &= \alpha \Pi +\beta^i\partial_i \varphi \label{eqn:dtphi} ~ , \\ 
\partial_t \Pi &= \alpha \gamma^{ij}\partial_i\partial_j \varphi +\alpha\left(K\Pi -\gamma^{ij}\Gamma^k_{ij}\partial_k \varphi - m^2 \varphi\right)\nonumber  \\
& + \partial_i \varphi \partial^i \alpha + \beta^i\partial_i \Pi \label{eqn:dtPi} ~ ,
\end{align}
where $\Pi$ is the conjugate momentum density, as defined by Eq.~\eqref{eqn:dtphi}, $K_{ij} = \frac{1}{2\alpha} \left( -\partial_t \gamma_{ij} + D_i\beta_j + D_j \beta_i \right)$ is the extrinsic curvature, and $K$ is its trace.

For all simulations, the initial conditions for the scalar field are set to a constant across the grid as $\operatorname{Re}[\varphi( t=0,r)]=\varphi_0$, $\operatorname{Im}[\varphi(t=0,r)]=0$, $\operatorname{Re}[\partial_t \varphi (t=0,r)] = 0$ and $\operatorname{Im}[\partial_t \varphi (t=0,r)] = \mu \varphi_0$. 
The initial amplitude $\varphi_0$ is arbitrary since we neglect the backreaction of the matter on the metric, but we use a value of order unity that can then be rescaled for different physical densities.
Here we have chosen $\varphi_0=0.1$.

The stress energy tensor for the complex scalar field is
\begin{equation}
\begin{aligned}
    T_{\mu\nu} &= \frac{1}{2}\left[\left(\nabla_\mu\varphi\nabla_\nu\varphi^* + \nabla_\nu\varphi\nabla_\mu\varphi^*\right)\right. \\
    &\left. - g_{\mu\nu}\left(\nabla^\lambda\varphi \nabla_\lambda\varphi^*\right)\right] - g_{\mu\nu} V\left(|\varphi|\right) ~.
\end{aligned}
\end{equation}
This can be expressed in terms of the energy, momentum and stress densities measured by the normal observers, as defined by the decomposition
\begin{equation}
    T_{\mu\nu} = \rho n_\mu n_\nu + S_\mu n_\nu + n_\mu S_\nu + S_{\mu\nu}\,,
\end{equation}
where
\begin{align}
    \rho &= n_\mu n_\nu T^{\mu\nu}\,,  \\
    S_i &= -\gamma_{i\mu}n_\nu T^{\mu\nu}\,, \\ 
    S_{ij} &= \gamma_{i\mu}\gamma_{j\nu}T^{\mu\nu}\,.
\end{align}

Since we do not include the backreaction of the scalar field onto the metric, or self-interactions, the results for a purely real scalar field can be obtained by simply considering the contributions of the real part of the field in the above expressions.

\subsection{Diagnostic quantities}\label{sec:diagnostic}

In the Newtonian description, the dynamical friction force in the $i$-direction $F_i$ is simply the integral of the gravitational force from each element of the fluid density acting on the BH, that is
\begin{equation}
    F_i = \int \frac{M \rho}{r^2}\frac{x^i}{r} ~ dV ~. \label{eqn:F_vol}
\end{equation}
In the regime where the matter reaches a stationary state, this will equal the net $i$-momentum flux of the fluid out of a closed surface around the volume:
\begin{equation}
    F_i = \int T_i^j dS_j ~. \label{eqn:F_flux}
\end{equation}
In the fully relativistic case the first expression is no longer valid but the second expression may be, provided that one adopts a coordinate system that approaches Minkowski spacetime sufficiently quickly at spatial infinity.\footnote{See Gourgoulhon \cite{Gourgoulhon:2007ue} for the exact requirements on the ADM quantities. Note that these are satisfied by the boosted isotropic coordinates we have used, but would not be, for example, in Cartesian Kerr-Schild coordinates.}
In this case we are quantifying approximately the change in the ADM momentum of the spacetime
\begin{equation}
    F_i = \partial_t P_i^{ADM} \eqsim \int T_i^j dS_j ~.
    \label{eq:Fsurfaceintegral}
\end{equation}
The approximation here comes from the fact that we are measuring at a large but finite radius (and not at spatial infinity, where the ADM momentum is actually defined). We are also neglecting the effect of the metric changing due to the presence of the fluid. Assuming that the fluid energy density is $O(\epsilon)$, the corresponding correction in the metric is $\delta g_{\mu\nu} \sim O(\epsilon)$, but this backreaction effect will only enter our force calculation at $O(\epsilon^2)$, firstly in neglecting the change in the flux calculated above as a result of $\delta g_{\mu\nu}$, and secondly by not accounting for any momentum loss from gravitational wave fluxes (which are $O(\delta g_{\mu\nu}^2)$). For small matter energy densities $\rho M^2 \ll 1$ the fixed background calculation will therefore be a good approximation. In superradiant clouds we expect $\rho M^2 \sim 10^{-5}$ in the most optimistic scenarios~\cite{Brito:2014wla}, and even lower densities are expected for accretion, making the approximation suitable for the intended applications.

However, as noted above, Eq.~\eqref{eqn:F_flux} can only be applied once the system has reached a stationary state. Before such a state is reached the momentum flux feeds both the momentum of the accumulating cloud, in addition to exerting a drag on the BH via $\delta g_{\mu\nu}$. What we are interested in measuring is only the latter of these two quantities. For our simulations, we therefore found it useful to derive a relativistic equivalent to the volume integral in Eq.~\eqref{eqn:F_vol}, since this converged to a constant value far earlier than the flux integral. Such an expression can be obtained by considering Gauss's law in the four-dimensional volume for the current $J^\mu = T^\mu_\nu \zeta^\nu$. Here $\zeta^\nu = \delta^\nu_\mu$ must be a Killing vector in the $\mu$ direction for the asymptotically flat space far from the BH (see \cite{Clough:2021qlv}), then the quantity we want to extract is
\begin{equation}
    {\cal S}_\mu = \int_{\Sigma} ~ \sqrt{g} ~ T^\rho_\nu \nabla_\rho \delta^\nu_\mu  ~ d^3x
\end{equation}
For the spatial direction $i$ we have $\zeta^\nu = \delta^\nu_i$ and
\begin{align}
    {\cal S}_i &= \int_{\Sigma} ~ \sqrt{g} ~ T^\mu_\nu ~^{(4)}\Gamma^\nu_{\mu i} ~ d^3x \\
    &= \int_{\Sigma} (- \rho \partial_i \alpha + S_j \partial_i \beta^j + \alpha S^k_j ~^{(3)}\Gamma^j_{ki}) ~ \sqrt{\gamma} d^3x ~.
\end{align}
Here we have used the fact that the components of $\zeta^\mu$ are constant in space, whilst in the second line we express the integrand in terms of the 3+1 ADM quantities. Whilst clearly the expression is coordinate-dependent within the volume, the asymptotic behavior guarantees that the volume integral is that measured by observers at infinity, such that the total four-vector $F_\mu$ transforms as expected under the Poincar\'e group (although note the comments in the previous section regarding the effect of our time-independent coordinate choice). 

To account for the presence of a singularity in the volume we must replace the full three-dimensional volume $\Sigma_{\rm out}$ with the sum of the volume outside some inner surface $\partial\Sigma_{\rm in}$ enclosing the singularity and the flux through that surface:

\begin{multline}
F_i = \underbrace{\int_{\Sigma_{\rm out} - \Sigma_{\rm in}} ~ \sqrt{g} ~ T^\mu_\nu ~^{(4)}\Gamma^\nu_{\mu i} ~ d^3x}_{{\rm dynamical ~ friction}, ~ {\cal S}_i} \\ + \underbrace{\int_{\partial\Sigma_{\rm in}} ~ \alpha T^j_i ~ dS_j}_{{\rm momentum ~ accretion},~ {\cal F}_{i,{\rm in}}}
\label{eqn:force-extr}
\end{multline}
Clearly the latter term should be identified with Bondi accretion of the fluid onto the BH, which also contributes to the drag force on the BH. We note that in the relativistic case the division of the force between accretion and dynamical friction is highly observer-dependent: consider for example an observer at infinity, who will not see any accretion onto the BH as the fluid will freeze at the horizon. As a result we measure the sum of both contributions, which is the most physically relevant quantity.
For verification, we also extract the surface integral of the momentum flux at the boundary of the spatial volume
\begin{equation}
    \mathcal{F}_i = \int_{\partial\Sigma_{out}} \alpha N_j T^j_i ~ dS 
    = \int_{\partial\Sigma_{out}} N_j ( \alpha \gamma^{jk} S_{ki} - \beta^j S_i ) ~ dS
\end{equation}
and the volume integral of the momentum density of the cloud
\begin{equation}
    \mathcal{Q}_i = \int \alpha T^0_i ~ dV = \int S_i ~ dV
\end{equation}
in order to check that these reconcile with $F_i$ in the expected way, and to demonstrate that the scalar cloud does at late times settle into a steady state such that the flux agrees with the volume integral.
We describe these tests in more detail in Appendix~\ref{app:code_validation}.

\subsection{Numerical methods}
\label{subsec:numerics}

We explore a range of four values of the scalar mass, $\alpha_{\rm s} = (0.05, 0.2, 0.5, 1.0)$, and for each measure the force for a range of velocities $v$, choosing (without loss of generality) the velocity to be in the $x$ direction.

We use an adapted version of the open source numerical relativity code $\textsc{GRChombo}$ \cite{GRChombo} to solve Eqs.~\eqref{eqn:dtphi} and \eqref{eqn:dtPi} on a fixed metric background in the boosted isotropic coordinates described above.
The scalar field is evolved by the method of lines with fourth-order finite difference stencils, Runge-Kutta time integration, and a hierarchy of grids with 2:1 resolution.
The value of the metric and its derivatives are calculated locally from the analytic expressions at each point.

We choose the size of the simulation domain $L$ such that the outer radius $r$ over which we integrate the flux will be larger than the de Broglie wavelength of the field, i.e. such that $\mu v r \gg 1$. This corresponds to the regime of validity for the nonrelativistic analytic results we compare to, and is most challenging to achieve numerically for cases of small $\alpha_{\rm s}$.
Therefore for the case of $\alpha_{\rm s}=0.05$ we choose $L=4096M$, for $\alpha_{\rm s}=0.2$ and $\alpha_{\rm s} = 0.5$ we use $L=2048M$, and for $\alpha_{\rm s}=1$, $L=1024M$.
We use $9$ $(2:1)$ refinement levels for $\alpha_{\rm s}=0.05$, $8$ for $\alpha_{\rm s}=0.2$ and $\alpha_{\rm s} = 0.5$, and $7$ for $\alpha_{\rm s}=1$,
with the coarsest level having $128^3$ grid points, although we exploit the quadrant symmetry of the problem in Cartesian coordinates to reduce the domain to $64^2 \times 128$ points.
This ensures that we maintain a spatial resolution of $dx = 0.0625M$ around the horizon of the BH, which is at $r \sim M/2$, in each case.

In order to resolve the temporal oscillations of the field sufficiently, we use a coarsest time step $dt_{\rm coarse}$ such that we have at least $32$ time steps per period of oscillation, i.e. $T = 2\pi/\mu > 32~ dt_{\rm coarse}$.

The form of the metric naturally imposes ingoing boundary conditions at the horizon, due to the causal structure of the BH.
We implement nonzero, time oscillating boundary conditions for the scalar field by setting the field to be spatially constant in the radial direction, extrapolating from the values within the numerical domain. This simulates the effects of a roughly constant energy density, but can introduce unphysical effects in very long simulations. These effects can be easily identified by varying the domain size, but ultimately limit the time for which the growth of the cloud can be studied. 
For all our setups we perform code validation and convergence tests, the details of which can be found in Appendix~\ref{app:code_validation}.


\section{Background on dynamical friction}
\label{sec:df_bg}

In this section we summarize previous results in dynamical friction that are relevant to this work, and formulate the relativistic approximations to which we will fit our numerical results.

The effect of an object losing momentum due to the gravitational interaction with the medium it moves through was first introduced by Chandrasekhar \cite{Chandrasekhar:1943ys}. 
The idea was formulated to describe the force exerted on a nonrelativistic star propagating through a gas of heavy noninteracting particles (other stars).
Since then Chandrasekhar's dynamical friction result has been applied to many other astrophysical problems, such as a relativistic projectile moving through a gaseous medium \cite{Syer:1994vr, Barausse:2007ph, Ostriker:1998fa, 1989ApJ...336..313P, 1971ApJ...165....1R, 1980ApJ...240...20R}, and to scalar dark matter in the nonrelativistic regime \cite{Hui:2016ltb, Bar_Or_2019,Berezhiani_2019,Lancaster_2020,Annulli:2020ilw,Annulli:2020lyc, Hartman:2020fbg}. 
The impact of such environments on the dynamics of compact objects and the potential for the resulting dephasing of GW signals to act as a probe of the matter has been considered in several works~\cite{Macedo:2013qea, Ferreira:2017pth, Hannuksela:2018izj, Zhang:2019eid, Eda:2013gg, Eda:2014kra, Kavanagh:2020cfn,  Yue:2019ozq, Bertone:2019irm, Hannuksela:2019vip, Edwards:2019tzf, Yue:2017iwc}. 

The classic Chandrasekhar result for a nonrelativistic perturber in a collisionless medium \cite{Chandrasekhar:1943ys} is
\begin{equation}
    F_{\rm Chandra} = 4 \pi \rho \left( \frac{M}{v} \right)^2 \ln \left(\frac{b_{\rm max}}{b_{\rm min}}\right)\,,
    \label{eqn:Chandra}
\end{equation}
where $b_{\rm min} \sim M/v^2$ is the size of the perturber or the capture impact parameter,
and $b_{\rm max}$ is the maximum impact parameter.
Relativistic corrections to this result were found in Ref.~\cite{Syer:1994vr} for the weak-scattering limit and in Ref.~\cite{1989ApJ...336..313P} for a collimated flow,
showing that the Chandrasekhar result was modified by the relativistic correction
\begin{equation}
    \text{Relativistic correction} = \gamma^2\,(1+v^2)^2\,, \label{eqn:relcorrection}
\end{equation}
as a multiplicative factor to the nonrelativistic Chandrasekhar result of Eq.~\eqref{eqn:Chandra}.
Here the factor of $(1+v^2)^2$ comes from the increase in the deflection angle of the fluid at relativistic velocities, $\alpha = 2M(1+v^2)/bv^2$ (notice that in the limit where $v=1$ we recover the factor $1+v^2=2$ that differentiates the deflection of photons from non relativistic particles). The factor of $\gamma^2$ accounts for the relativistic momentum of the fluid as seen in the frame of the perturber. This is $\tilde{\rho} \gamma v$, taking $\tilde{\rho}$ to be the asymptotic energy density in the perturber frame. Noting that $\rho= \tilde{\rho}/\gamma$ is the energy density in the rest frame of the fluid gives the second $\gamma$ factor.

In supersonic collisional fluids (with Mach number $\mathcal{M} = v/c_s > 1$, where $c_s$ is the sound speed of the fluid), a further correction is required such that the rest mass density $\rho$ is replaced by the sum of the density and the (isotropic) pressure of the fluid $p$~\cite{1989ApJ...336..313P,1971ApJ...165....1R, 1980ApJ...240...20R, Ostriker:1998fa}. That is, we introduce another multiplicative factor of
\begin{equation}
    \text{Pressure correction} = \frac{\rho + p }{\rho}
\end{equation}
to Eq.~\eqref{eqn:Chandra}, in addition to the relativistic correction in Eq.~\eqref{eqn:relcorrection}.
For collisional fluids in the subsonic case, Ostriker~\cite{Ostriker:1998fa} worked out the Newtonian case and Barausse~\cite{Barausse:2007ph} extended the calculation to the relativistic case, but these results will not be relevant to the present study, since we expect to always be in the supersonic regime, as discussed below. 
Another interesting correction comes from finite size effects of the surrounding fluid, as studied in \cite{Vicente:2019ilr, Annulli:2020lyc}. These give a nontrivial correction to the results that assume an infinite homogeneous reservoir, but such effects are neglected in this work.

The analytic expression for the dynamical friction force on a point-like object generated by a scalar field cloud, derived in Ref.~\cite{Hui:2016ltb}, relates the force to the relative velocity between the object and the fluid $v$, the Compton wavelength of the scalar $\lambda_{\rm C} = 1/\mu$, the BH mass $M$ and the size of the perturbed region $r$.\footnote{The value of $r$ is dependent on the environment and the past history of the BH passage through the fluid. The spatial extent of the surrounding fluid provides an upper limit on $r$, but in the case where the BH enters a cloud of fluid and progresses through it, $r$ will initially be zero and then increase over time as the wake behind it forms, such that $r \sim vt$. This time dependence in $r$ was captured by the dynamical calculations of Ref. \cite{Annulli:2020lyc}.}
It has the form
\begin{equation}
    F_{\rm d, nonrel} = 4 \pi \rho  \left( \frac{M}{v} \right)^2 \left\{ \ln{(2kr)} - 1 - {\rm Re} [\Psi(1+i\beta)] \right\}\,,
    \label{eqn:Fd_nonrel}
\end{equation}
where $k = \mu v$, $\beta = \alpha_{\rm s} / v$ is the ratio between the impact parameter where deflection becomes significant and accretion dominates ($b_{\rm min} \sim M/v^2$) and the de Broglie wavelength of the scalar ($\lambda_{\rm dB} = 1/k$), and $\Psi$ is the digamma function.\footnote{\url{https://mathworld.wolfram.com/DigammaFunction.html}}
Depending on the ratio $\beta = \alpha_{\rm s} / v$ this takes the form
\begin{align}
    &F_{\rm d}|_{\beta\to 0} \sim \frac{4\pi\rho}{v^2}\,\bigg[\log(2kr ) - 1 + 0.577 - 1.202 \beta^2\bigg]\label{eqn:Fd_sbeta}\,,\\
    &F_{\rm d}|_{\beta\to \infty} \sim \frac{4\pi\rho}{v^2}\,\left[\log(\frac{2v^2r}{M} ) - 1 - \frac{1}{12\beta^2}\right]\label{eqn:Fd_lbeta}\,,
\end{align}
where again $k = \mu v$, $\rho$ is the asymptotic value of the scalar field rest mass density, and $r$ is the size of the perturbed volume -- normally the size of the dark matter cloud in which the perturber is immersed. 
(This can now be identified with the maximum impact parameter in the standard Chandrasekhar result $b_{\rm max} \sim 2r$ in the limit of large $\beta$ \cite{Hui:2016ltb}.)

These expressions are valid only in the case where $r$ is larger than the de Broglie wavelength of the field, $kr\gg 1$. For small $kr$, Ref.~\cite{Hui:2016ltb} provides an alternative expression which is in principle valid for any radius $r$ but only for the case of smaller $\beta$, which is
\begin{equation}
        F_{\rm d}|_{\beta < 1} = 4 \pi \rho  \left( \frac{M}{v} \right)^2 \left[ {\rm Cin}(2kr) + \frac{\sin{(2kr)}}{2kr} - 1 \right]\,,
    \label{eqn:Fd_nonrel_smallkr}
\end{equation}
where ${\rm Cin}(z) = \int_0^z (1 - \cos(t)) dt /t$ is the cosine integral. 
In this work we fit the regime described by Eq.~\eqref{eqn:Fd_nonrel}, which allows us to cover a wider range of scalar masses. We leave a full approximation for the small $kr$ regime to future work.

\begin{figure}[t]
	\centering
	\includegraphics[width=0.45\textwidth]{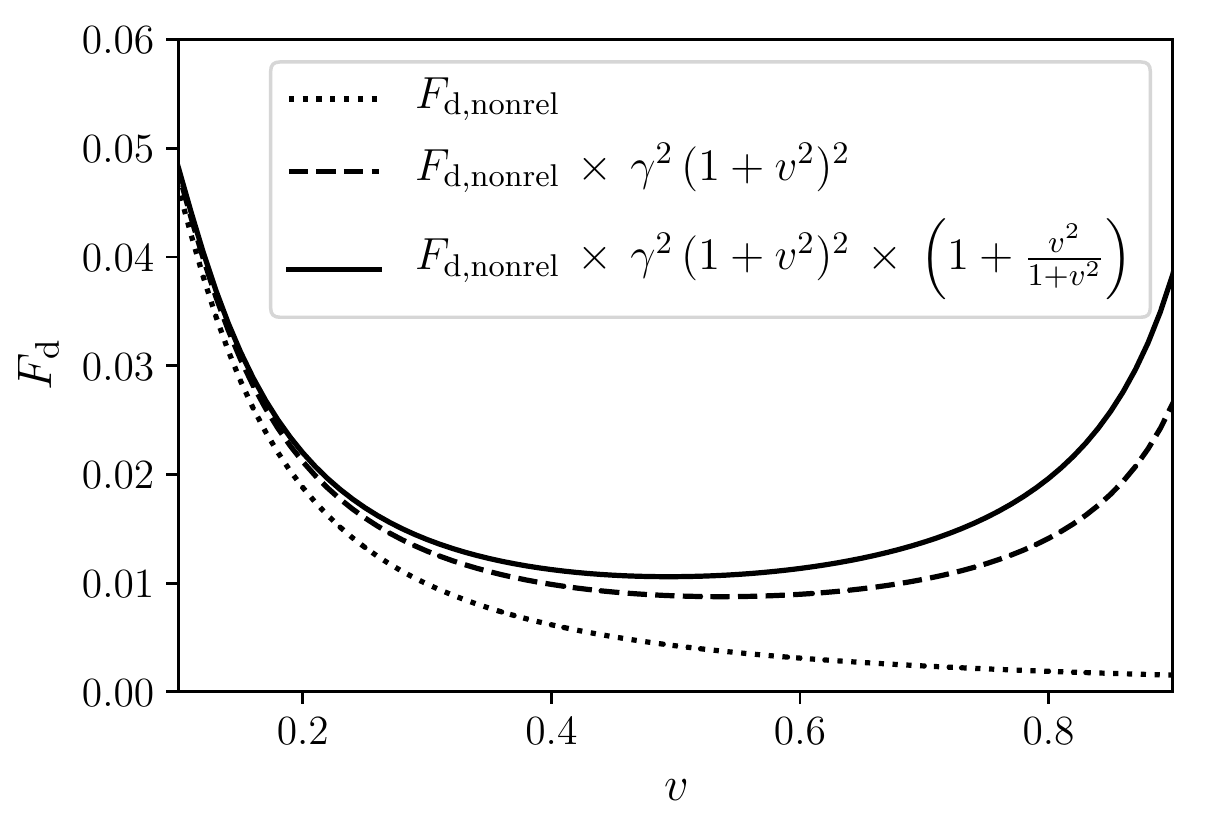}
	\caption{Analytic approximation for the nonrelativistic case for a scalar field cloud (dotted line) and with a relativistic correction term (dashed line) and a further correction for a pressure-like term, as for supersonic collisional fluids (solid line).}
	\label{fig:analytic}
\end{figure}
Note that dynamical friction is distinct from the drag force created by Bondi accretion \cite{Bondi:1952ni, Bondi:1944jm}, i.e. accretion of the fluid momentum onto the BH as it moves through the cloud 
\begin{equation}
    F_{\rm Bondi} = 4\pi \lambda M^2 \rho v (v^2 + c_s^2)^{-3/2}\,,
    \label{eqn:Bondi}
\end{equation}
with $\lambda$ an $O(1)$ number. 
In the small-$\alpha_{\rm s}$ regime, Bondi accretion is negligible in comparison to the effect of dynamical friction. This is the case for our results for $\alpha_{\rm s} = 0.05$.
However, we find that for larger masses of the scalar field its contribution becomes nonnegligible, and we must include it in order to better fit our simulation results.
Physically, this is because the surface integral calculated in \cite{Hui:2016ltb} assumes a point mass perturber. This neglects several finite size effects, one of which is accretion. In the low-mass scalar case that was considered there, the impact of accretion is small because the incoming waves are mostly scattered rather than absorbed. In the particle regime, and at lower velocities where deflection is more significant, accretion can no longer be neglected.

To apply this to our relativistic scenario, we start with the nonrelativistic result of Eq.~\eqref{eqn:Fd_nonrel}, and take as an ansatz the relativistic correction required for collisional fluids as above,
\begin{equation}
    \text{Relativistic correction} = \gamma^2\,(1+v^2)^2\,,
\end{equation}
assuming as before that it enters as a multiplicative factor. 
\begin{figure*}[t]
	\centering
	\includegraphics[width=0.9\textwidth]{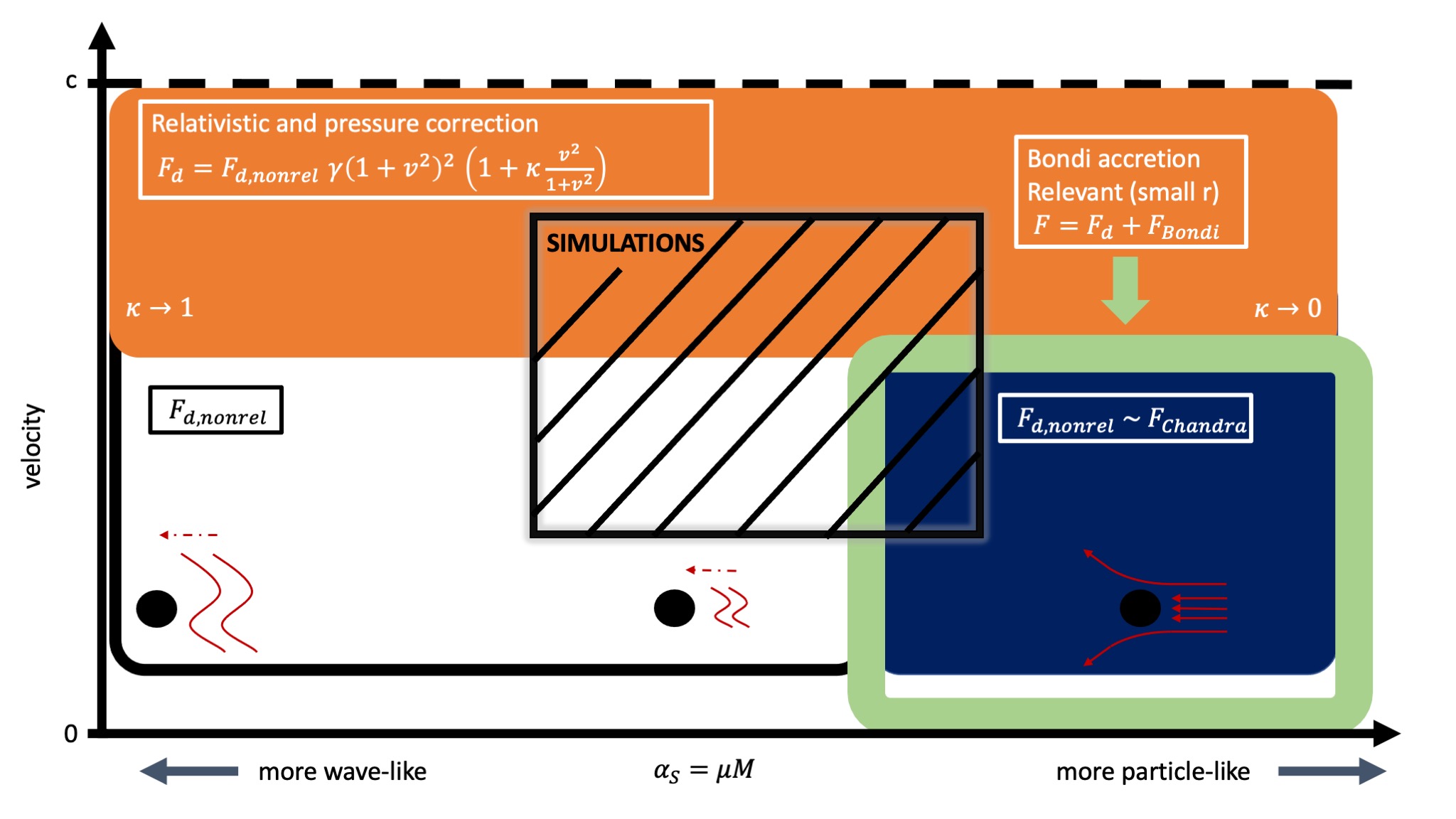}
	\caption{
    The diagram shows schematically the regime covered by our simulations in the $\alpha_S = \mu M$ - velocity plane.
    We consider a range of intermediate masses which cover the wave regime of $\alpha_S =0.05$ up to $\alpha_S = 1$, where wave effects begin to turn off on the scale of the BH horizon and accretion onto the BH becomes non negligible. We study velocities $0.2 < v < 0.8$ at which relativistic corrections become significant. The expression for $F_d$ describes the dynamical friction, with the $\alpha_S$-dependent parameter $\kappa$ describing pressure effects fitted to our numerical results. See the main text for further details.
    }
	\label{fig:velocity_vs_alpha}
\end{figure*}
We also assume that the rest-mass density $\rho$ will be replaced with the sum of the density and pressure $\rho+p$. Whilst an inhomogeneous scalar field does not have a well defined equation of state, we can consider a plane wave solution, which is what the BH will ``see'' in the asymptotically flat region, in its rest frame,
\begin{equation}
    \varphi(x,t) = \varphi_0 ~ e^{i(\omega t - kx)}\,,
\end{equation}
where $\omega^2 = \mu^2 + k^2$ and $k=\mu v$. We can then obtain approximate expressions for $\rho$ and $p$ from the stress-energy tensor as 
\begin{align}
    &\rho = \frac{1}{2} \dot\varphi \dot\varphi^* + \frac{1}{2}(\partial_i \varphi \partial_i \varphi^*) + V
         = \mu^2 |\varphi|^2 (1 + v^2)\,, \\
    &p_x = \frac{1}{2} \dot\varphi \dot\varphi^* + \frac{1}{2} (\partial_x \varphi \partial_x \varphi^*) - V
         = v^2 \mu^2 |\varphi|^2 \,. \\
    &p = \frac{1}{2} \dot\varphi \dot\varphi^* - \frac{1}{6} (\partial_i \varphi \partial_i \varphi^*) - V
         = \frac{1}{3}v^2 \mu^2 |\varphi|^2  \,. 
\label{eqn:rho_p}
\end{align}
Note that since the fluid is no longer isotropic, we could consider the pressure correction as either $\rho + p$, where $p$ is the averaged pressure in all three spatial directions, or as $\rho + p_x$, where $p_x$ is the pressure in the direction of the BH motion. Either choice will be an approximation, but we find the latter to be a better fit and more physically intuitive -- the pressure in the $x$ direction is what plays the greatest role in resisting the BH motion.\footnote{In particular, we find that the value of the parameter $\kappa$ that we introduce below, tends to a value $\sim 1$ with this choice at lower scalar masses, which is the expected behavior.}

Using this, we estimate the pressure correction as
\begin{equation}
    \text{Pressure correction} = \frac{\rho + p_x }{\rho} = 1 + \frac{v^2}{(1+v^2)}\,.
\label{eqn:pressure_corr}
\end{equation}
Note that this also justifies the assumption that for our setup we are always in the supersonic regime, since we have $c_s \sim v^2$ and thus a Mach number $\mathcal{M} \sim 1/v$ greater than 1.

The expected effect of such relativistic corrections on the result of \cite{Hui:2016ltb} is illustrated in Fig.~\ref{fig:analytic}. 
The final expression that we use to fit our results has the form
\begin{equation}
     F_{\rm d} = F_{\rm d, nonrel}\times\,\gamma^2(1+v^2)^2\,\times\,\left(1+\kappa\frac{v^2}{1+v^2}\right)\,,
     \label{eqn:Fd_rel}
\end{equation}
where $\kappa$ is a factor we add to control the strength of the pressure-like correction.
We include this factor as we expect this correction to become weaker (or completely turn off) as we increase the mass of the scalar field, where it would start to behave more like a particle than a wave.
As we will show, this approximation turns out to be a very good fit to our results for all values of the mass, with the lower masses corresponding to $\kappa\sim1$.

A diagram illustrating schematically the different scalar mass and velocity regimes, and placing our simulations within them, is given in Fig.~\ref{fig:velocity_vs_alpha}.

\section{Results}
\label{sec:results}
In this section we present the results we obtain for the dynamical friction force and its scaling with velocity derived from our simulations. 
We compare these results to the expected analytic expression, as detailed in Sec.~\ref{sec:df_bg}, and fit the value of $\kappa(\alpha_{\rm s})$ for each value of $\alpha_{\rm s}$.

As explained in Sec.~\ref{subsec:numerics} above, we consider four different values of $\alpha_{\rm s}$, and for each case calculate the force for velocities in the range $v\in (0.3,0.9)$, except in the case of $\alpha_{\rm s}=1$, where we add an extra point at $v=0.2$ to better explore the effect of Bondi accretion. 

The different scalar masses we choose, $\alpha_{\rm s} = (0.05,0.2,0.5,1)$, probe the two different regimes of validity of the expression in Eq.~\eqref{eqn:Fd_nonrel}.
In particular, as detailed in Appendix~\ref{app:digamma_approx}, the smallest mass we consider ($\alpha_{\rm s} = 0.05$) is consistent with the expression in Eq.~\eqref{eqn:Fd_sbeta}, valid for $\beta\ll1$. 
The cases of $\alpha_{\rm s} = 0.5$ and $1$ probe the large-$\beta$ regime of Eq.~\eqref{eqn:Fd_lbeta}.
We also chose a point from the an intermediate mass range ($\alpha_{\rm s} = 0.2$), where neither of the approximations is valid and one needs to consider the exact expression in Eq.~\eqref{eqn:Fd_nonrel}.

\begin{figure}[t!]
	\centering
	\includegraphics[width=0.45\textwidth]{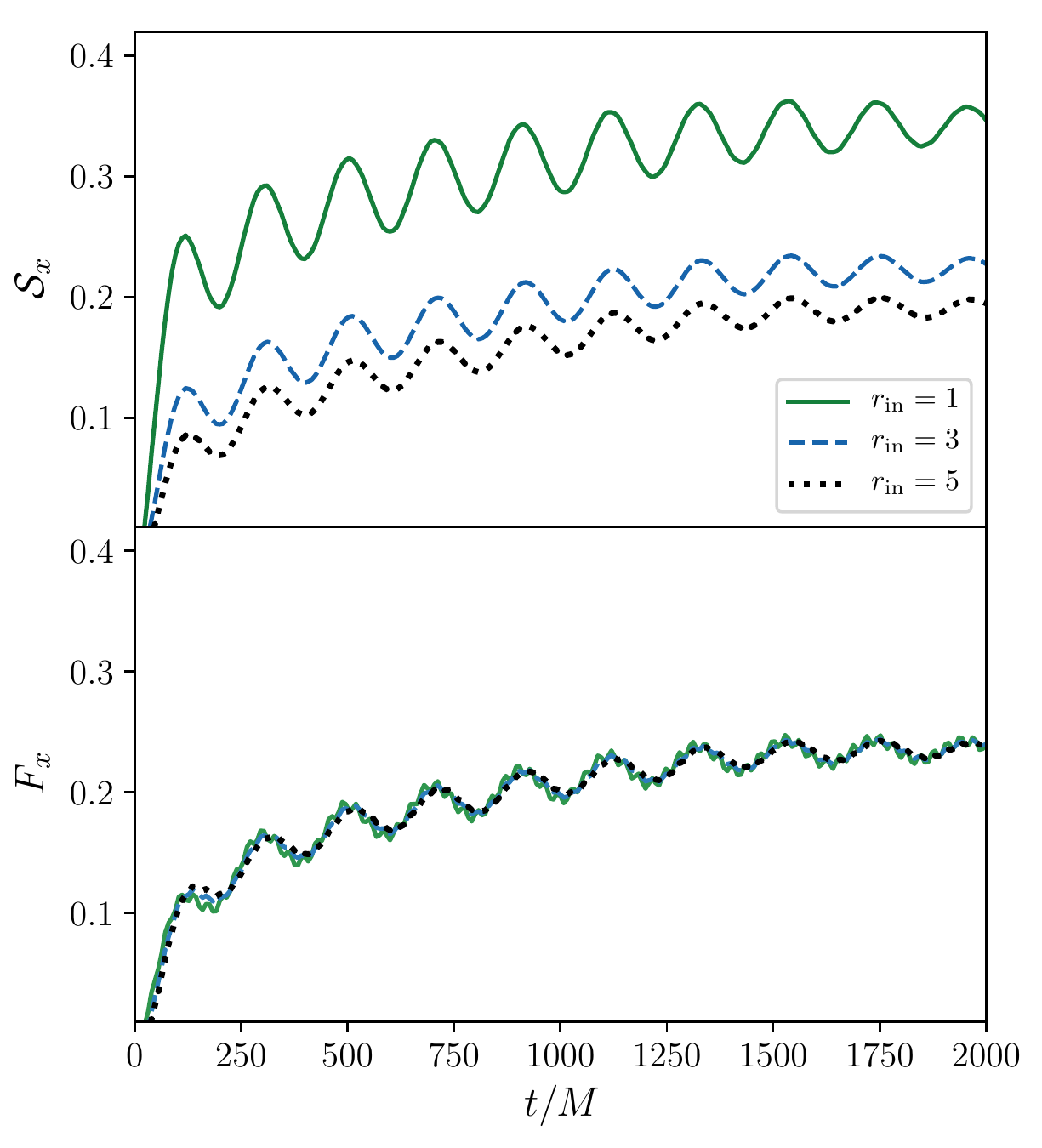}
	\caption{Dynamical friction force as a function of time extracted from one of our simulations. Top panel: contribution to the total force coming only from the ``dynamical friction'' term ${\cal S}_x$ in Eq.~\eqref{eqn:force-extr}. This is calculated excising a volume of three different radii $r_{\rm in}$ around the BH: $r_{\rm in}=1$ (green solid line), $r_{\rm in}=3$ (blue dashed line) and  $r_{\rm in}=5$ (black dotted line). 
	Bottom panel: total force $F_x$ in Eq.~\eqref{eqn:force-extr}, including the contribution ${\cal F}_{x,\,{\rm in}}$ from the flux through the inner surface $\Sigma_{\rm in}$ containing the BH. Including both terms, the total force is independent of the value of $r_{\rm in}$.}
	\label{fig:source+flux}
\end{figure}
\begin{table}[t]
\begin{center}
\begin{tabular}{|p{0.5cm}||p{0.2cm}|p{0.4cm}|p{0.4cm}|p{1cm}|}
 \hline
    $\alpha_{\rm s}$ & $r_{\rm in}$ & $r_{\rm out}$ & $L$ & $T_{\rm final}$\\
 \hline
 $0.05$     & $5$   & $900$     & $4096$    & $\gtrsim 5000$\\
 \hline
 $0.2$      & $5$   & $700$     & $2048$    & $\gtrsim 2500$\\
 \hline
 $0.5$      & $5$   & $600$     & $2048$    & $\gtrsim 2000$\\
 \hline
 $1$        & $10$  & $300$     & $1024$    & $\gtrsim 1000$\\
 \hline
\end{tabular}
\caption{Parameters that vary in our simulations depending on the parameter $\alpha_{\rm s}$, which compares the Compton wavelength of the scalar to the BH size. 
In the first two columns are the the inner and outer extraction radii for the force, $r_{\rm in}$ and $r_{\rm out}$, followed by the simulation box size, $L$, and the final time (at which we extract the diagnostic quantities) $T_{\rm final}$. These are all in units of the BH mass $M$, which we set to $1$ in our simulations.}
\label{tab:sim_pars}
\end{center}
\end{table}
\begin{figure*}[t!]
	\centering
	\includegraphics[width=0.9\textwidth]{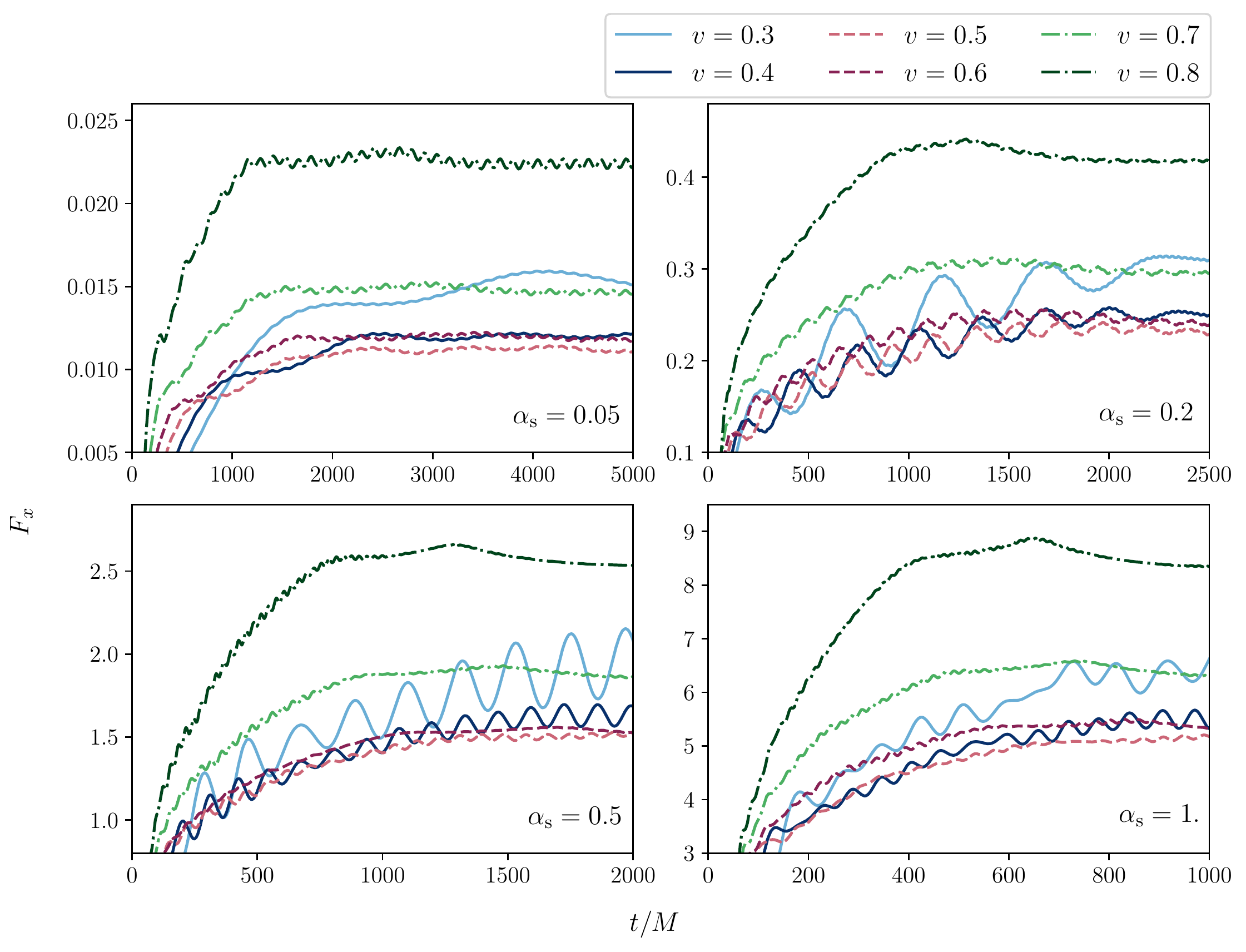}\\
	\caption{Time evolution of the total force $F_x$ extracted from the simulations [Eq.~\eqref{eqn:force-extr}]. 
	The four blocks show the the different scalar field masses we consider: $\alpha_{\rm s}=0.05$ (top left), $\alpha_{\rm s}=0.2$ (top right), $\alpha_{\rm s}=0.5$ (bottom left) and $\alpha_{\rm s}=1$ (bottom right).
	For each case we plot the evolution of the field for the range of velocities $v=0.3-0.8$.
	In lighter and darker solid blue are $v=0.3$ and $0.4$, dashed red (light to dark) -- $v=0.5$ and $0.6$ and dash-dotted green (light to dark) are $v=0.7$ and $0.8$.}
	\label{fig:Force_vs_T}
\end{figure*}
\begin{figure*}[ht]
	\centering
	\includegraphics[width=0.9\textwidth]{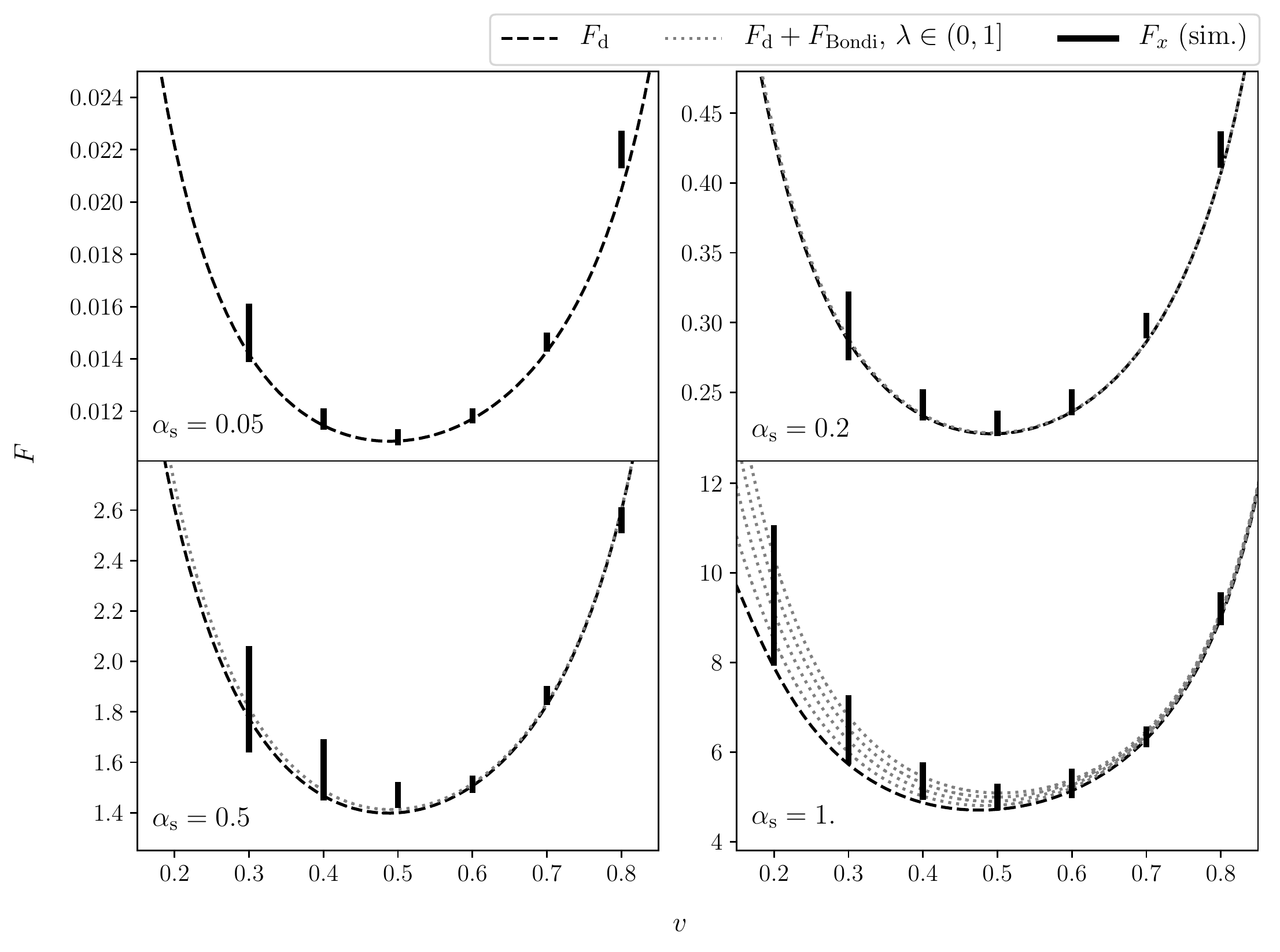}
	\caption{
	Scaling of the total force, $F$, on the BH due to the scalar field with relative velocity $v$.
	The four blocks again represent this result for the four mass cases we consider.
	The dashed lines in these plots show the analytic expression for the dynamical friction force $F_{\rm d}$ including the relativistic corrections, Eq.~\eqref{eqn:Fd_rel}; the dotted lines show the analytic approximation of the total force on the BH, including both the contributions from dynamical friction and Bondi accretion, $F_{\rm Bondi}$. We include a range of values for the order one parameter $\lambda$ in Eq.~\eqref{eqn:Bondi}. The error bars (solid black) represent the total force we extract from the simulations, $F_x$ [Eq.~\eqref{eqn:force-extr}], with the size of the error bars coming mainly from the amplitude of the oscillations of the force, as seen in Fig~\ref{fig:Force_vs_T}. Further details are given in Appendix~\ref{app:code_validation}.
	}
	\label{fig:Force_vs_v}
\end{figure*}

As outlined in Sec.~\ref{sec:diagnostic}, we extract the quantities in Eq.~\eqref{eqn:force-extr}.
For the radius $r$ of the volume $\Sigma_{\rm out}$ we take an outer extraction radius $r = r_{\rm out}$ large enough for the analytic approximation that we want to compare with [Eq.~\eqref{eqn:Fd_nonrel}] to be valid: $r_{\rm out} \gg 1/(\mu v)$. Therefore $r_{\rm out}$ varies depending on the value of $\alpha_{\rm s}$.\footnote{Note that at larger radii the field takes longer to settle into a stationary profile and the force takes much longer to reach a constant value. The competition between this timescale and the accumulation of noise from the outer boundaries puts an upper limit on the outer extraction radius, $r_{\rm out}$ at which we can extract the force.}
We choose the inner extraction radius such that the the BH horizon is fully enclosed in this region, $r_{\rm in} \gtrsim r_{\rm hor}$ (in isotropic Schwarzschild coordinates, $r_{\rm hor} = 0.5$).
In principle we could use the same value for all our simulations, but we found that the flux is noisy when $r_{\rm in}$ is chosen to be too close to the horizon, and we require larger radii for the larger $\alpha_{\rm s}$.
Recall that the inner radius is used to measure the flux ${\cal F}_{x,\,{\rm in}}$, which we can identify with the Bondi accretion contribution to the drag force. 

For the cases where the flux is nonnegligible compared to ${\cal S}_x$, the choice of $r_{\rm in}$ determines the split between Bondi accretion drag and dynamical friction. Such a split is inherently gauge dependent, and so we measure the sum of the dynamical friction and accretion forces, which does not vary according to the choice of $r_{\rm in}$.
We demonstrate this in Fig.~\ref{fig:source+flux}, where we see that although the ``dynamical friction'' term ${\cal S}_x$ in Eq.~\eqref{eqn:force-extr} can differ significantly depending on the choice of $r_{\rm in}$ (this is shown on the top panel), once we add the contribution ${\cal F}_{x,\,{\rm in}}$ from the flux through the inner surface the total force is the same regardless of the choice of $r_{\rm in}$.

In Table~\ref{tab:sim_pars} we give the different extraction radii we have used to obtain our results from the simulations for the four different mass cases.
We also show the box size $L$ for each of these cases and the approximate time $T_{\rm final}$ at which we extract the diagnostic quantities.
The larger simulation times and box sizes for the smaller velocities come from the fact that a scalar field with a longer wavelength compared to the BH light-crossing time will take longer (in units of M) to settle into a stationary profile.

The time evolution for the force is shown in Fig~\ref{fig:Force_vs_T}. 
The four panels present the four different scalar mass cases, and for each case we show the evolution for the range of velocities we consider.
We find that in all cases the force grows initially and settles into a roughly constant value after some time, with some remaining oscillations, which tend to be larger for smaller velocities.
As the velocity decreases the overall amplitude of the force gets smaller in the regime $v\sim 0.3 - 0.5$, after which this pattern shifts and the force starts to grow again quite rapidly.
Note that the time scale for the four different plots is different, since the higher the mass of the field, the faster it settles into a stationary profile and a constant value for the force.
The approximate simulation times for each case are given in the $T_{\rm final}$ column of Table~\ref{tab:sim_pars}. However, note that due to the field taking longer to settle in the lower-velocity cases (both into a constant value and a lower amplitude of oscillation), we needed to run some cases longer before extracting a final value for the force.
Values on the $y$ axis are also not shown since they are different for all four cases and vary depending on the radius of the sphere we are considering; what is important is the overall trend in each case.
Our goal is not to obtain a specific value for the force in each case, but rather to fit the dependence on $v$ in our expression, which can then be applied at all radii $r$, provided that $r\gg(1/\mu v)$.
Note that we have confirmed that different radii give consistent results for the scaling of the force with velocity.

Fig.~\ref{fig:Force_vs_v} illustrates the scaling of the force with velocity for each of the scalar masses, $\alpha_{\rm s}$.
The dashed lines represent the analytic expression for the dynamical friction force $F_{\rm d}$, and the dotted lines represent the total force on the BH due to the combined effect of dynamical fiction and Bondi accretion, $F=F_{\rm d}+F_{\rm Bondi}$.
The error bars are the force $F_x$ that we extract from our simulations.
We estimate the size of the error bars from the amplitude of the field oscillations at the time of extraction, plus the errors coming from finite resolution as identified by our convergence tests (discussed in more detail in Appendix~\ref{app:code_validation}).

\begin{figure}[ht]
	\centering
	\includegraphics[width=0.45\textwidth]{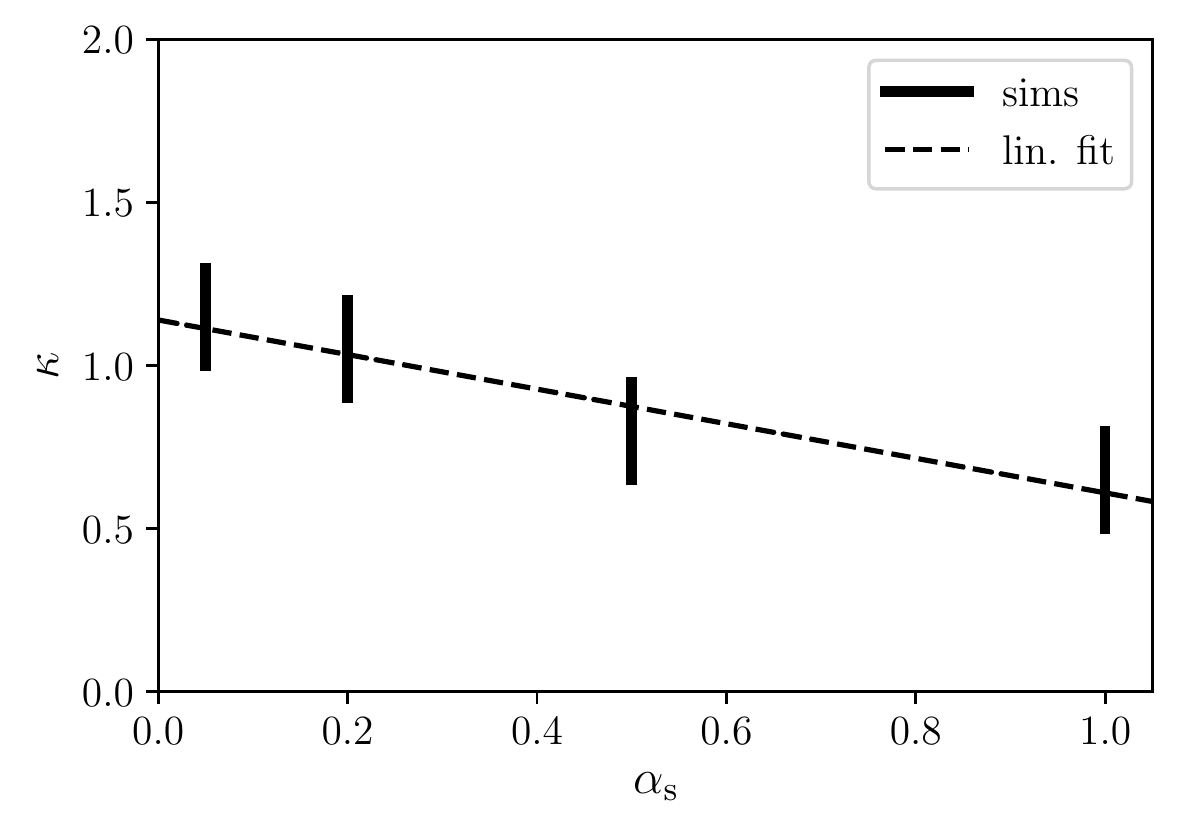}
	\caption{This plot shows the scaling of the pressure correction factor $\kappa$ with scalar mass, $\alpha_{\rm s}$.
	The error bars are derived from our simulations (see Fig.~\ref{fig:Force_vs_v}).
	We fit a straight line though these, with equation $\kappa = 1.14-0.53\alpha_{\rm s}$ (dashed line).
	Note that this expression is only valid in the range of scalar masses we consider, $\alpha_{\rm s}\in(0.05,1)$. 
	We would expect that $\kappa$ has a value $\sim 1$ for smaller $\alpha_{\rm s}$ and tends to zero for higher $\alpha_{\rm s}$, and we see that this is consistent with our result.}
    \label{fig:kappa_vs_alphas}
\end{figure}

We find that Eq.~(\ref{eqn:Fd_rel}) gives a good fit to the results in each case (dashed line in the plots on Fig.~\ref{fig:Force_vs_v}).
As expected, we find that low-mass cases fit well with the pre-factor for the pressure-like correction $\kappa\sim1$, and for higher masses $\kappa$ starts to decrease.
The values of $\kappa$ that give us the best fit to our results (with their estimated errors)
for the different scalar mass cases, as well as an approximate linear fit for these is shown in Fig.~\ref{fig:kappa_vs_alphas}.
As the field gets heavier, it starts to behave less like a wave and more like a particle, so it is not surprising that in the critical mass regime (where the wavelength of the field becomes comparable to the size of the BH, $\alpha_{\rm s}\sim 1$), this pressure-like correction derived from a plane-wave solution would weaken, and the field will start to behave more like a pressureless fluid.
We find that $\kappa$ decreases from $\kappa \sim 1$ for $\alpha_{\rm s} \sim 0.05$ to $\kappa \sim 0.5$ for $\alpha_{\rm s} \sim 1.0$ and can expect that for even larger field masses, this pressure correction will fully disappear.

As discussed above, for all masses except for $\alpha_{\rm s}=0.05$, the momentum accretion onto the BH starts to have a significant contribution the total force that we measure. 
This can be associated with Bondi accretion, discussed in Sec.~\ref{sec:df_bg} [Eq.\eqref{eqn:Bondi}], and so we cannot assume that its effect is negligible compared to dynamical friction.
The dotted lines in Fig.~\ref{fig:Force_vs_v} represent the expected total force onto the black hole, including this contribution, so that $F = F_{\rm d} + F_{\rm Bondi}$.
In the case of $\alpha_{\rm s}=0.2$ (top right) accretion is negligible, and this line fully overlaps with the dashed black line showing $F_{\rm d}$ alone. However in the cases of $\alpha_{\rm s}=0.5$ and $\alpha_{\rm s}=1$ (bottom left and right) we find that the effect is significant at lower velocities.
For $\alpha_{\rm s}=1$, especially by taking the additional point at $v=0.2$, where the difference between the two lines becomes larger, we can show that the line including Bondi accretion (with $\lambda\sim0.5$) is most consistent with our results.

\section{Discussion}
\label{Discussion}

In this paper we have numerically studied the dynamical friction force $F_{\rm d}$ on a BH immersed in a uniform-density region of scalar fluid of extent $r$ as a function of the relativistic boost velocity $v$. We have demonstrated that the scaling of the total drag force experienced by the BH goes as
\begin{equation}
\begin{aligned}
    F =& F_{\rm d} + F_{\rm Bondi} \\
    =& F_{d, \rm nonrel}\times\,\gamma^2(1+v^2)^2\,\times\,\left(1+\kappa\, \frac{v^2}{1+v^2}\right) \\
    &+ 4\pi \lambda M^2 \rho v (v^2 + c_s^2)^{-3/2}\,,
\end{aligned}
\end{equation}
where we find $\kappa= \kappa(\alpha_{\rm s}) \in (0,1)$, $\lambda\in(0,1)$ and $c_s = \kappa\, v^2/(1+v^2)$.\footnote{Note that our results are not sensitive to the exact value of $c_s$ in the expression for the Bondi accretion, and we find consistency for any $c_s\in(0,0.5)$. The exact form we quote here is simply taken to match the form of $p_x/\rho$ in the dynamical friction expression.}
This is consistent with the case of a supersonic collisional fluid, combining a relativistic correction of $\gamma^2\,(1+v^2)^2$ with the assumption that the (now velocity-dependent) pressure correction of the scalar is well described by the approximate relation
\begin{equation}
    \rho + p_x \sim \rho \left(1+\kappa\, \frac{v^2}{1+v^2}\right) ~.
\end{equation}
We find that the constant $\kappa$ depends on the scalar mass, decreasing from $\kappa \sim 1$ for $\alpha_{\rm s} \sim 0.05$ to $\kappa \sim 0.5$ for $\alpha_{\rm s} \sim 1.0$. We expect it to approach zero in the limit of large $\alpha_{\rm s}$, thus recovering a result consistent with collisionless particles. An approximate linear fit for the intermediate mass regime that we study is provided in Fig. \ref{fig:kappa_vs_alphas}.

Our result can be applied to studies of EMRI dephasing in the final orbits around a SMBH with a superradiant or accretion-driven cloud. However, some care will need to be given to the value of $r$ that is chosen to represent the size of the cloud that influences the BH. As we have shown, the cloud takes time to develop dynamically, and therefore at each point in the orbit, the size of the cloud will depend on both the local extent of the scalar density, and also on the inspiral history. As shown in Ref. \cite{Annulli:2020lyc} in the nonrelativistic case, this can change the result significantly compared to the naive picture of an infinite uniform density environment. In addition, the effective pressure of the scalar fluid depends on the scalar mass, and this will determine the extent to which the cloud recovers to its original configuration over successive orbits - that is, one should consider whether the cloud is significantly depleted by the Bondi accretion, or alternatively whether its local density is enhanced on subsequent orbits. The full problem of simulating a relativistic BH orbiting a larger one immersed in a superradiant scalar cloud profile remains open, with the main challenge being the difference in the timescales involved.

The results in this work are derived for the case of a complex scalar field in order to avoid having to deal with the high frequency amplitude oscillations expected for a real scalar field.
Nevertheless, we confirm for a few of the cases that the real field oscillates with the expected frequency of $\omega = 2\alpha_{\rm s}$, and the average value of these oscillations is consistent with the result we obtain for the complex field. We provide more detail on this in Appendix~\ref{app:real_field}. Whilst the average effect over time for a real scalar would be consistent with the complex case, in principle the impact of the higher frequency oscillations on the trajectory of the perturber is another distinctive signature of this type of scalar matter.

In addition to providing specific results for scalar dark matter, the framework that we have described and validated provides the groundwork for the study of a wider range of bosonic candidates, and could be simply extended to include nontrivial self-interactions such as those of axions (as in \cite{Helfer:2016ljl}) or dark photons through the implementation of a massive vector (Proca) field. One could also study the impact of adding spin to the BH perturber, or a curved trajectory in an asymptotically curved spacetime (as in \cite{Barausse:2007ph}), by updating the metric background used.

\section*{Acknowledgements}
\vspace{-0.2in}
\noindent We thank the GRChombo collaboration (www.grchombo.org) for their support and code development work. We thank J. Bamber, V. Cardoso, R. Croft, M. Radia, J. C. Aurrekoetxea and H. Witek for helpful conversations.
DT, KC and PGF acknowledge funding from the European Research Council (ERC) under the European Unions Horizon 2020 research and innovation programme (grant agreement No 693024). 
E.B. and T.H. are supported by NSF Grants No. PHY-1912550 and AST-2006538, NASA ATP Grants No. 17-ATP17-0225 and 19-ATP19-0051, NSF-XSEDE Grant No. PHY-090003, and NSF Grant PHY-20043.
LH acknowledges support by the Department of Energy DE-SC011941 and a Simons Fellowship in Theoretical Physics.
The simulations presented in this paper used the Glamdring cluster, Astrophysics, Oxford, DiRAC resources under the projects ACSP218 and ACTP238 and PRACE resources under Grant Numbers 2020225359 and 2018194669 and computational resources at the Maryland Advanced Research Computing Center (MARCC). This work was performed using the Cambridge Service for Data Driven Discovery (CSD3), part of which is operated by the University of Cambridge Research Computing on behalf of the STFC DiRAC HPC Facility (www.dirac.ac.uk). The DiRAC component of CSD3 was funded by BEIS capital funding via STFC capital grants ST/P002307/1 and ST/R002452/1 and STFC operations grant ST/R00689X/1. In addition used the DiRAC at Durham facility managed by the Institute for Computational Cosmology on behalf of the STFC DiRAC HPC Facility (www.dirac.ac.uk). The equipment was funded by BEIS capital funding via STFC capital grants ST/P002293/1 and ST/R002371/1, Durham University and STFC operations grant ST/R000832/1. DiRAC is part of the National e-Infrastructure. The PRACE resources used were the GCS Supercomputer JUWELS at J\"ulich Supercomputing Centre(JCS) through the John von Neumann Institute for Computing (NIC), funded by the Gauss Centre for Supercomputing e.V. (\url{www.gauss-centre.eu}) and computer resources at SuperMUCNG, with technical support provided by the Leibniz Supercomputing Center. The authors also acknowledge the Texas Advanced Computing Center (TACC) at The University of Texas at Austin for providing HPC resources that have contributed to the research results reported within this paper. URL: http://www.tacc.utexas.edu \cite{10.1145/3311790.3396656}.

\appendix

\section{Code validation}
\label{app:code_validation}

In this Appendix we discuss how we validate our code results, and provide some additional details on the fixed metric background used. 

\subsection{Code validation and coordinate choice}

As discussed in the main text, we evolve the field on a fixed background metric in boosted isotropic Schwarzschild coordinates.

The metric is validated by checking that the numerically calculated Hamiltonian and momentum constraints converge to zero with increasing resolution, as do the time derivatives of the metric components, i.e. $\partial_t \gamma_{ij} = \partial_t K_{ij} = 0$ (calculated using the ADM expressions). This ensures that (ignoring the backreaction) the metric that is implemented is indeed stationary in the chosen gauge, consistent with it being fixed over the field evolution. 

The advantage of using isotropic coordinates is that the asymptotic behavior is appropriate for measuring an approximate ADM momentum for the system at large $r$, due to the rate at which the quantities approach flat Minkowski spacetime asymptotically. The downside is that since the time coordinate corresponds to that of the asymptotic observers, the lapse goes to zero and time freezes around the horizon.\footnote{The implementation of these coordinates is improved by the use of an analytic continuation of the lapse in which its value becomes negative within the horizon, as described in \cite{Bamber:2020bpu}.}
\begin{figure}[t]
	\centering
	\includegraphics[width=0.45\textwidth]{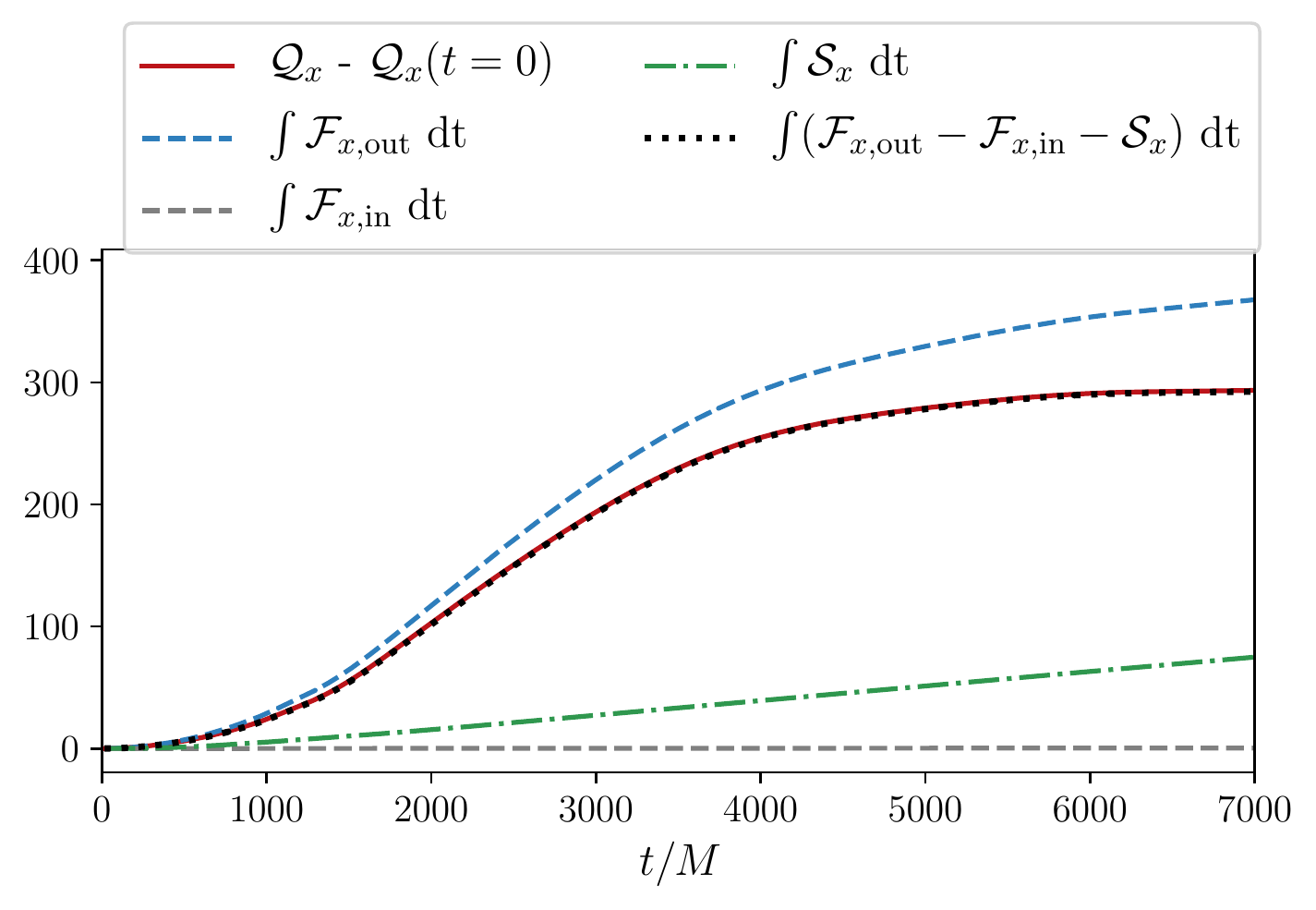}
	\caption{Agreement between the time-integrated quantities defined in Eq.~\eqref{eqn:conserve}. The grey and blue dashed lines show the momentum flux through the inner and outer extraction surfaces, respectively, the green dash-dotted line is the ``source'' within the volume between those two surfaces, the red solid line is the momentum density in the $x$-direction and the black dotted line is the RHS of Eq.~\eqref{eqn:conserve}. We see that the momentum of the cloud settles into a roughly stationary value at around $t\sim5000$ and that there is good agreement between the surface flux and volume integrals.
	}
	\label{fig:Integrals}
\end{figure}
This means that any ingoing waves tend to ``bunch up'' there. Given sufficient resolution outside the outer horizon, the ingoing nature of the metric prevents the errors this introduces from propagating into the region far from the BH, and unresolved waves are effectively damped away by grid precision close to the horizon. Provided we are not interested in extracting quantities very close to the horizon, these coordinates work in practice, as shown below by the conservation of the flux and volume integrals shown in Fig.~\ref{fig:Integrals}, where we confirm the expected relation (see \cite{Clough:2021qlv})
\begin{align}
	\partial_t \left( \int_{\Sigma} d^3 x \sqrt{\gamma} ~ \mathcal{Q}_i \right) =& - \int_{\partial\Sigma_{\rm out}} d^2 x \sqrt{\sigma} ~ \mathcal{F}_i\\
	&+ \int_{\partial\Sigma_{\rm in}} d^2 x \sqrt{\sigma} ~ \mathcal{F}_i + \int_\Sigma d^3 x \sqrt{\gamma} ~ \mathcal{S}_i\,,
	\label{eqn:conserve}
\end{align}
where we define $\partial\Sigma_{\rm in}$ to be the 2D surface cutting out the three-dimensional volume around the BH $\Sigma_{\rm in}$ (at $r_{\rm in}$), $\partial\Sigma_{\rm out}$ -- the 2D surface enclosing the volume of influence on the BH $\Sigma_{\rm out}$ (at $r_{\rm out}$), and $\Sigma = \Sigma_{\rm out} - \Sigma_{\rm in}$.
The ``charge'' $\mathcal{Q}_i$ is the momentum density in the $i$ direction defined by the coordinate basis vector $\zeta^{\mu} = \delta^\mu_i$, i.e.,
\begin{equation}
	\mathcal{Q}_i = n_\mu \zeta^\nu T^\mu_\nu = - \alpha T^0_i = - S_i\,,
\end{equation}
the ``flux'' $\mathcal{F}$ is the $i$-stress in the normal direction (aka the $i$-momentum flux out of the closed surface $\partial\Sigma$)
\begin{equation}
    \mathcal{F}_i = \alpha N_j J_i^j = \alpha N_j T^j_i = N_j ( \alpha \gamma^{jk} S_{ik} - \beta^j S_i )\,,
\end{equation}
and the reconcilling ``source'' term $\mathcal{S}$ is
\begin{align}
    \mathcal{S}_i &= \alpha T^\mu_\nu \nabla_\mu \zeta^\nu = \alpha T^\mu_\nu ~^{(4)}\Gamma^\nu_{\mu i} \\
    &= - \rho \partial_i \alpha + S_j \partial_i \beta^j + \alpha S^k_j ~^{(3)}\Gamma^j_{ik} ~.
\end{align}
Here $N_i$ is defined to be in the direction of the covector $s_i = \partial_i r$ (so $s_i= x^i/r$ in Cartesian coordinates) and it is normalized such that $\gamma_{ij} N^i N^j = 1$, whilst $n_\mu = (-\alpha,0,0,0)$ is the normal to the ADM spatial hypersurfaces.

\begin{figure}[t]
	\centering
	\includegraphics[width=0.45\textwidth]{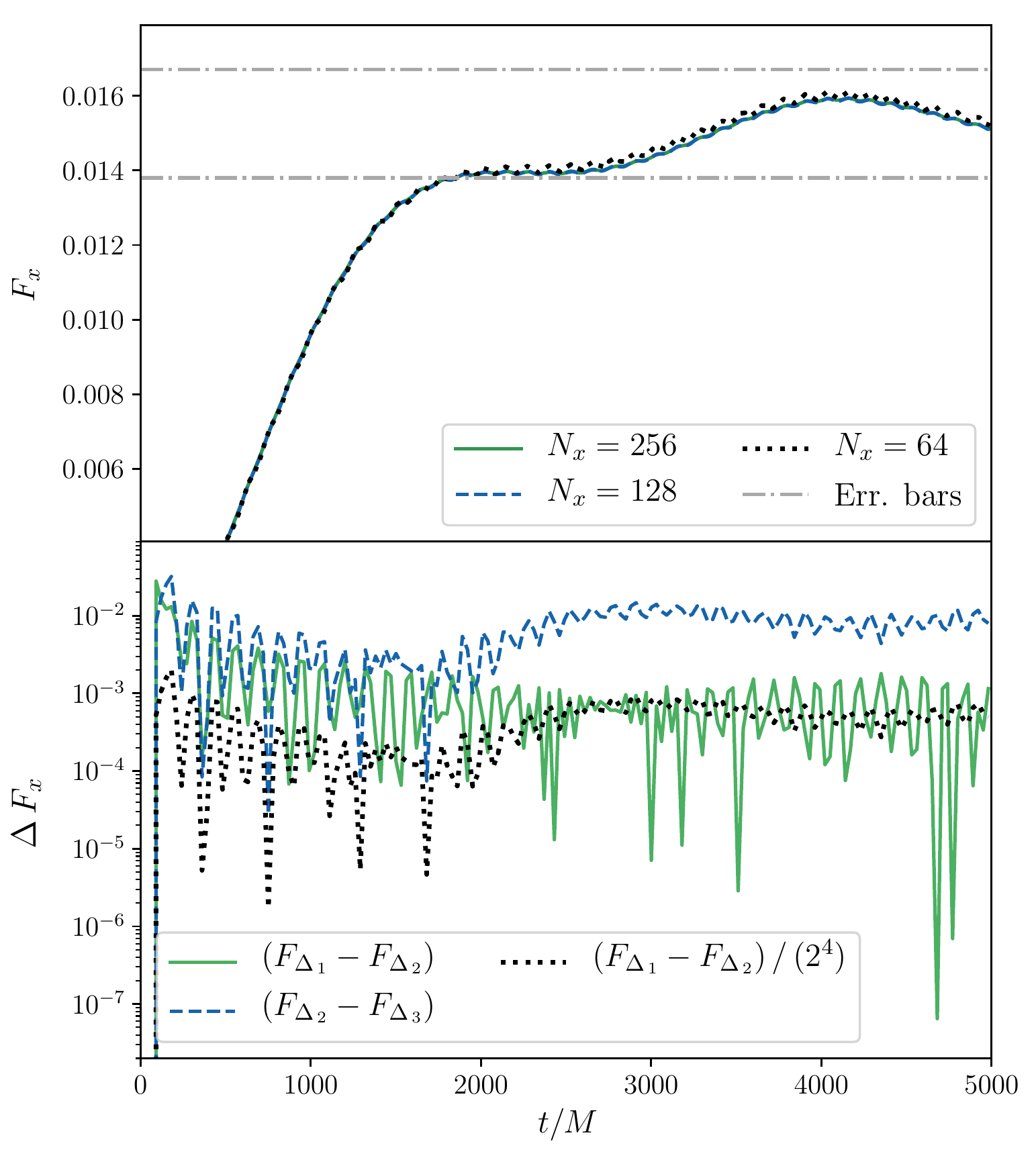}
	\caption{Top panel: evolution of the force extracted from the simulations for $\mu=0.05$ and $v=0.3$ at three different resolutions. We show the the high resolution case ($N_x=256$, solid green line), the medium resolution  used for our results ($N_x=128$, dashed blue line), and low resolution case ($N_x = 64$, dotted black). The grey dash-dotted lines show how we extract the error bars (in Fig.~\ref{fig:Force_vs_v}) from this simulation.
	Bottom panel: relative difference between the solutions at the high/low resolutions and the medium one in solid green and dashed blue, respectively. We also show the difference in the solutions between the low and medium resolutions divided by a convergence factor of $2^4$ (black dotted line), showing that we achieve fourth-order convergence.}
	\label{fig:convergence}
\end{figure}
\subsection{Convergence testing and error estimation}

We have performed convergence tests for each of the masses considered at $v=0.3$ and $v=0.8$.
For each of these we have computed our results at three resolutions, $\Delta_1$, $\Delta_2$ and $\Delta_3$.
For the cases $\mu=0.05,\,0.2$ and $0.5$, the high resolution, $\Delta_1$ is for $N_x=256$, the medium resolution $\Delta_2$ (that we run all our simulations at) corresponds to $N_x = 128$, and the low resolution corresponds to $N_x=64$.
On the top of Fig.~\ref{fig:convergence} we show as an example the evolution of the force for the three resolutions from one of our simulations.
The low resolution case is slightly more noisy than the medium and high resolution cases, but they are consistent and the difference is well within our error bars, the measurement of which is illustrated with the grey dash-dotted lines. These error bars from the oscillation of the force are the ones presented in Sec.~\ref{sec:results}.

\begin{figure}[t]
	\centering
	\includegraphics[width=0.45\textwidth]{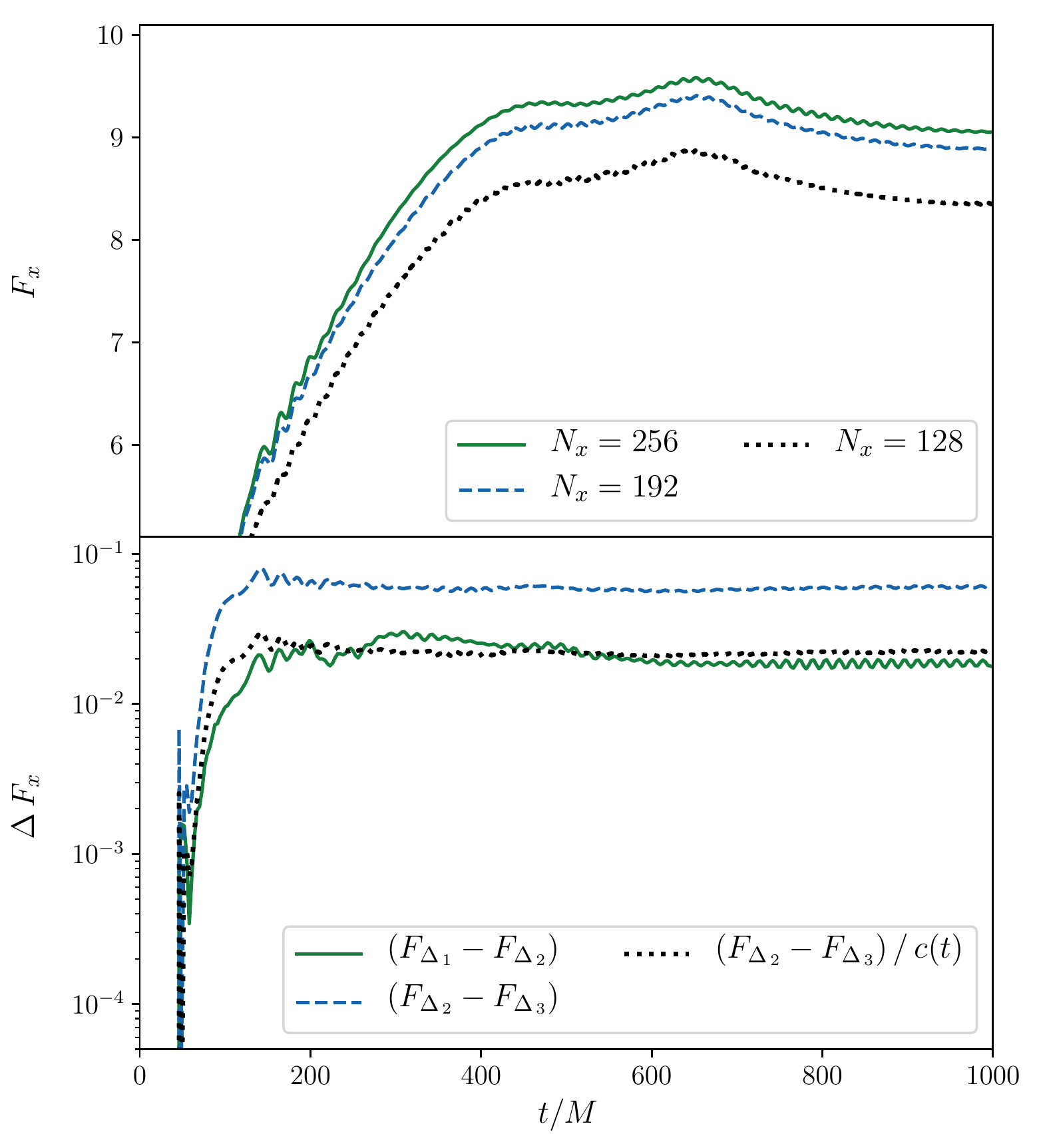}
	\caption{Top panel: evolution of the force extracted from the simulations for $\mu=1$ and $v=0.8$ at four different resolutions, where the green solid line shows the high-resolution case ($N_x = 256$), the dashed blue line corresponds to $N_x=192$ and the black dotted line corresponds to the resolution used in all of our simulations ($N_x=128$).
	Bottom panel: relative differences in the force between the $\Delta_1$ and $\Delta_2$ resolutions (solid green line) and between the $\Delta_2$ and $\Delta_3$ resolutions (dashed blue line). The dotted black line is the difference in the solution between $\Delta_2$ and $\Delta_3$ resolutions divided by the convergence factor $c(t)$, as calculated by Eq.~\eqref{eqn:conv_factor}, showing fourth-order convergence.}
	\label{fig:convergence2}
\end{figure}

The convergence factor is defined as the ratio of the relative differences between the solution at the medium and the low/high resolutions,
\begin{equation}
    c(t) = \frac{||F_{\Delta_1} - F_{\Delta_2}||}{||F_{\Delta_2} - F_{\Delta_3}||}\,.
\end{equation}
In the case where the three resolutions we have considered have number of grid points $N_x=(64,128,256)$ (so that $\Delta_1/\Delta_2=\Delta_2/\Delta_3=2$), in the limit $\Delta\to 0$ the convergence factor is expected to behave as,
\begin{equation}
    \lim_{\Delta\to0}c(t) = 2^n\,,
\end{equation}
where $n$ is the order of the finite difference scheme used to evolve our initial configuration.
We have $n = 4$, so that the expected convergence factor is $c(t) = 2^4$.
We show the results of this test in the bottom panel of Fig.~\ref{fig:convergence}, where the green solid line shows the difference in the solutions for the high and medium resolution cases, and the blue dashed line is the difference between the low and medium resolution values of the force. 
The black dotted line is the the difference between the high and medium resolution cases divided by the expected convergence factor for this case, $2^4$, lies on top of the solid green line, showing that indeed we get fourth-order convergence for the simulation.

The case of ($\mu=1$, $v=0.8$) was the most challenging from the set of simulations presented.
The evolution of the force, $F_x$, extracted from this simulation is shown on the top panel of Fig.~\ref{fig:convergence2}.
In this case the low resolution, $N_x=64$, was found to be outside of the convergence regime, so instead we calculated the force at resolutions $N_x = (128,196,256)$. 
For this set of resolutions we no longer have $\Delta_1/\Delta_2=\Delta_2/\Delta_3=2$ and the convergence factor in the continuum limit is given by 
\begin{equation}
    \lim_{\Delta\to0}c(t) = \frac{\Delta_1^n - \Delta_2^n}{\Delta_2^n - \Delta_3^n} \sim 2.7\,,
    \label{eqn:conv_factor}
\end{equation}
for $N_x = 128,196,256$ and $L=1024$, so that $\Delta_1 = 0.25$, $\Delta_2 = 0.1875$ and $\Delta_3 = 0.125$, and $n=4$.
The bottom panel of Fig.~\ref{fig:convergence2} shows that in this case we still achieve fourth-order convergence, with the low resolution now being $N_x=128$.

Although we confirm that the results are converging with resolution in the expected way, the resolution errors are comparable with those computed from the oscillations.
We can estimate the error on the solution at the highest resolution grid using the relation
\begin{equation}
    \epsilon_{\Delta_1}\sim\frac{1}{(\Delta_1/\Delta_2)^n-1}(F_{\Delta_1} - F_{\Delta_2})\sim0.08\,.  
\end{equation}
We apply this error to the highest resolution result. Note that if we were to use the lowest resolution for our results ($N_x=128$) the error would have been larger, $\epsilon_{\Delta_3}\sim 0.7$ (around $9\%$).
Similarly we have checked the error on the results at $N_x=128$ for $v=0.7$ (as that is expected to be lower) and we find that to be around $5\%$.
We also use this error as a conservative estimate of the resolution error in the extracted force from the simulations of $v=0.5,0.6$, $\alpha_{\rm s}=1$, using it to increase the error bars presented.
In a few other cases, we also identify and add convergence errors to the error bars in a similar way, but these are much less significant.

\section{Large and small \texorpdfstring{$\beta$}{Lg} approximations}\label{app:digamma_approx}
\begin{figure}[h]
	\centering
	\includegraphics[width=0.49\textwidth]{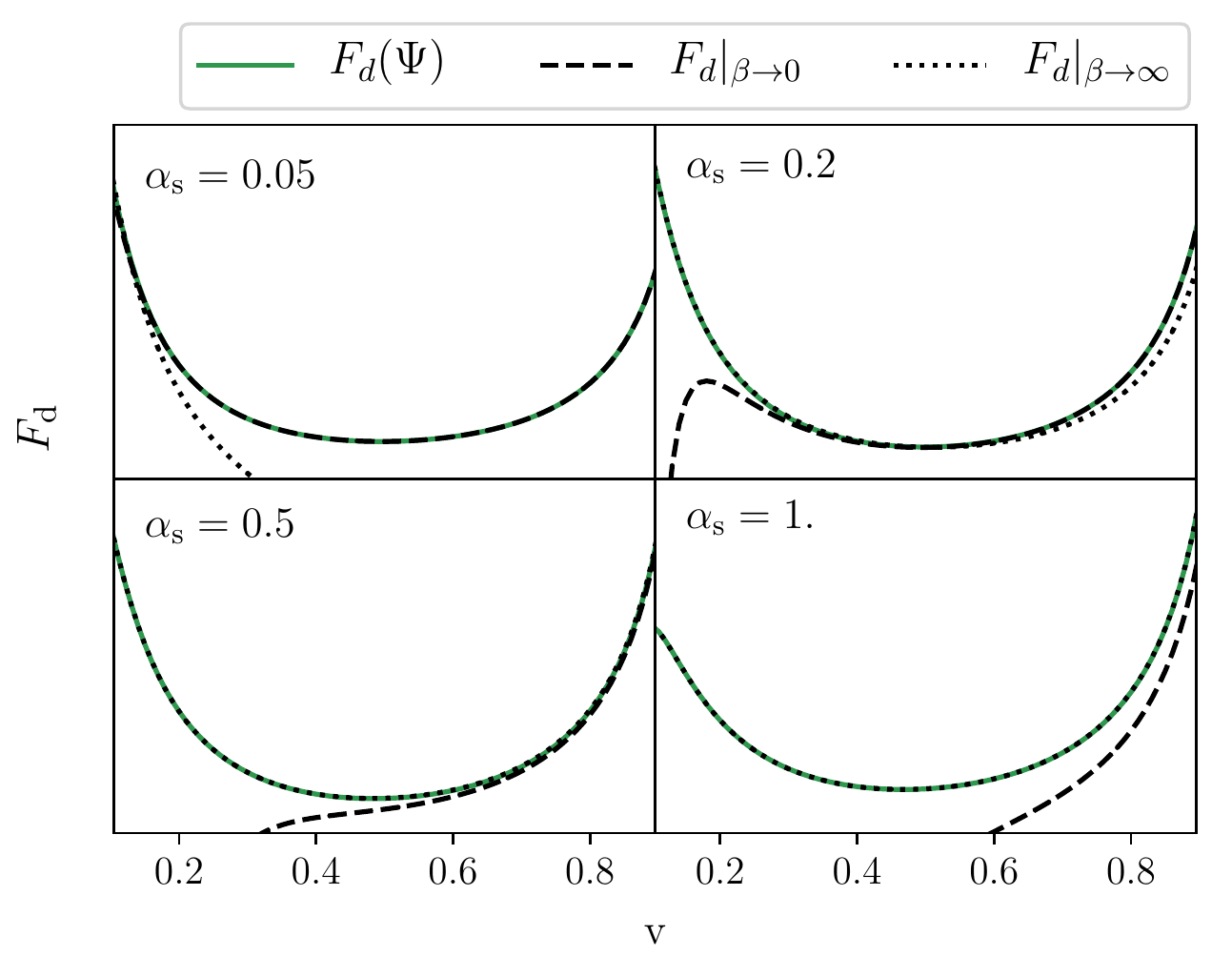}
	\caption{The expressions in Eqs.~\eqref{eqn:Fd_nonrel}--\eqref{eqn:Fd_lbeta} as a function of the relative velocity $v$ of the BH with respect to the scalar field, for the four values of $\alpha_{\rm s}$ that we simulate. The solid green lines show the exact expression in terms of the digamma function, while the dashed and dotted black lines correspond to the small-$\beta$ and large-$\beta$ approximations, respectively.}
	\label{fig:digamma_vs_beta}
\end{figure}
As discussed in Sec.~\ref{sec:df_bg}, the analytic approximation for the dynamical friction force on a BH from a scalar cloud (for a BH moving at nonrelativistic velocities), given in Eq.~\eqref{eqn:Fd_nonrel}, depends on the digamma function $\Psi(1+i\beta)$.
Although $\Psi$ can be evaluated and plotted with many computational packages such as Mathematica or Python, its shape is not immediately obvious, so it is useful to look at its approximate form in terms of $\beta = \alpha_{\rm s} / v$.
This can be done in the two regimes, where $\beta$ is either very small or very large:
\begin{align}
    \Re\left[\Psi(1+i\beta)\right] &\sim - 0.577 + 1.202 \beta^2 \quad (\beta\to 0)\,,\\
    \Re\left[\Psi(1+i\beta)\right] &\sim \ln\beta - \frac{1}{12\beta^2} \quad (\beta\to \infty)\,,
\end{align}
which leads to the approximate expressions in Eqs.~\eqref{eqn:Fd_sbeta} and \eqref{eqn:Fd_lbeta}.

To illustrate the range of validity of these approximate expressions, in Fig.~\ref{fig:digamma_vs_beta} we plot the scaling of the force $F_{\rm d}$ with velocity, given the exact expression containing the digamma function and the two approximations for small and large $\beta$.
As expected for the case of $\alpha_{\rm s} = 0.05$ (top left) the valid expression is the $\beta\to 0$ approximation. 
At large masses of the scalar, $\alpha_{\rm s} = 0.5$ and $1$, the solutions for $\beta\to \infty$ match the exact solution very well.
In the intermediate-mass case ($\alpha_{\rm s} = 0.2$), both expression match the exact solution in the range $v\sim0.4-0.5$, but the small-$\beta$ approximation fails at small velocities and the large-$\beta$ approximation is inaccurate at large $v$.

\section{Real scalar field}\label{app:real_field}
\begin{figure}[ht]
	\centering
	\includegraphics[width=0.45\textwidth]{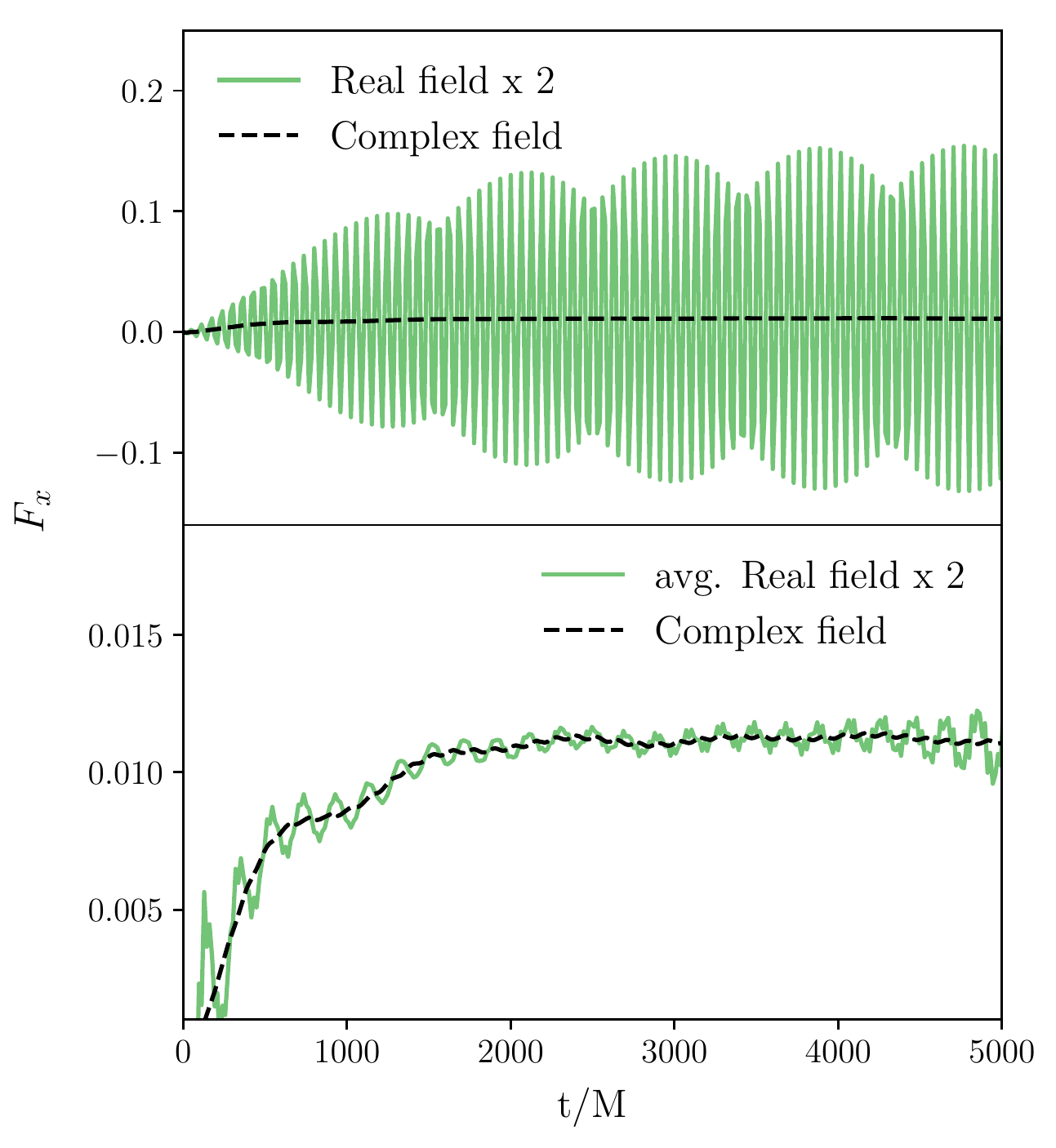}
	\caption{Time evolution of the force in the case of a real (solid green lines) and complex (dashed black lines) scalar field. The top panel shows the force as extracted from the simulation, whereas in the bottom panel we have removed the high-frequency oscillations, so as to see more clearly the agreement between the two cases.}
	\label{fig:real_field}
\end{figure}
A uniform real scalar field has a density that oscillates with a frequency related to its mass. This gives rise to a dynamical friction force that oscillates between positive and negative values over time.
This can make the results more difficult to interpret and lead to large error bars on the numerical results.
To avoid this issue we have performed all our simulations for a complex scalar field instead, as this can provide an asymptotically constant density that is twice the average of the real field density.
To confirm that this is indeed the case, we evolve the real scalar field in the case of $\alpha_{\rm s} = 0.05$ and show that the force in the complex field case is twice the average of that for the real field, as expected.

Fig.~\ref{fig:real_field} shows the evolution of the force as a function of time for $\alpha_{\rm s} = 0.05$ and $v=0.5$ in the case of a real scalar field (solid green lines) and the complex one (dashed black lines).
In the top panel, the real field force is as extracted from the simulations.
As expected for the real field, the force oscillates with very high frequency and amplitude.
We confirm that the frequency of the oscillations is related to the mass of the field as $\omega=2\alpha_{\rm s}$.
To confirm that the complex field gives the average of the oscillations in the real case we have removed the high-frequency oscillations using a filtered Fourier transform of the signal, as shown in the bottom panel of Fig.~\ref{fig:real_field}.

We observe that in the case of a purely real field, the dynamical friction force oscillates rapidly in time. Whilst the average effect over time would be consistent with the complex case, it should in principle be possible to observe the impact of the higher frequency oscillations on the trajectory of the perturber.

\bibliography{DynamicalFriction}

\end{document}